\documentclass[11pt,draftclsnofoot,journal,onecolumn]{IEEEtran}

\IEEEoverridecommandlockouts

\usepackage[dvips]{graphicx}
\usepackage{subfigure}
\usepackage{amsmath}
\usepackage{amssymb}
\usepackage{verbatim}
\usepackage{url}
\usepackage{cite}

\renewcommand{\IEEEQED}{\IEEEQEDopen}

\begin{document}

\title{Analysis Framework for Opportunistic Spectrum {OFDMA} and its Application to the IEEE~802.22 Standard}
\author{Jihoon Park, Przemys{\l}aw Pawe{\l}czak, P\r{a}l Gr{\o}nsund, and Danijela \v{C}abri{\'c}
\thanks{Jihoon Park, Przemys{\l}aw Pawe{\l}czak and Danijela \v{C}abri{\'c} are with the Department of Electrical Engineering, University of California, Los Angeles, 56-125B Engineering IV Building, Los Angeles, CA 90095-1594, USA (email: \{jpark, przemek, danijela\}@ee.ucla.edu).}
\thanks{P\r{a}l Gr{\o}nsund is with the Telenor, Snar{\o}yveien 30, 1331 Fornebu, Norway; Simula Research Laboratory, Martin Linges Vei 17, 1364 Fornebu, Norway; and the Department of Informatics, University of Oslo, Gaustadallen 23, 0373 Oslo, Norway (email: paalrgr@ifi.uio.no).}
\thanks{Part of this work has been published in the proceedings of IEEE GLOBECOM, Dec. 6--10, 2010, Miami, FL, USA~\cite{park_submitted_2010_gc}.}
\thanks{Also available at http://arxiv.org/abs/1007.5080.}}

\maketitle

\begin{abstract}
We present an analytical model that enables throughput evaluation of Opportunistic Spectrum Orthogonal Frequency Division Multiple Access (OS-OFDMA) networks. The core feature of the model, based on a discrete time Markov chain, is the consideration of different channel and subchannel allocation strategies under different Primary and Secondary user types, traffic and priority levels. The analytical model also assesses the impact of different spectrum sensing strategies on the throughput of OS-OFDMA network. The analysis applies to the IEEE 802.22 standard, to evaluate the impact of two-stage spectrum sensing strategy and varying temporal activity of wireless microphones on the IEEE 802.22 throughput. Our study suggests that OS-OFDMA with subchannel notching and channel bonding could provide almost ten times higher throughput compared with the design without those options, when the activity and density of wireless microphones is very high. Furthermore, we confirm that OS-OFDMA implementation without subchannel notching, used in the IEEE 802.22, is able to support real-time and non-real-time quality of service classes, provided that wireless microphones temporal activity is moderate (with approximately one wireless microphone per 3,000 inhabitants with light urban population density and short duty cycles). Finally, two-stage spectrum sensing option improves OS-OFDMA throughput, provided that the length of spectrum sensing at every stage is optimized using our model.
\end{abstract}

\IEEEpeerreviewmaketitle

\section{Introduction}
\label{sec:introduction}

One of the ways to combat artificial spectrum scarcity~\cite{staple_spectrum_2004} is to augment existing radio access techniques with Opportunistic Spectrum Access~\cite{Zhao_sigprocmag_2007} (OSA). Wireless networks with OSA capabilities are able to search for unused licensed portions of the radio spectrum and communicate over those vacant radio frequencies whenever available radio capacity is insufficient, while meeting the required interference constraints. 

Orthogonal Frequency Division Multiple Access (OFDMA) is a multiple access technique where orthogonally-divided frequency subcarriers are assigned to individual users of the network. A subcarrier assignment is usually performed by a central entity, often Base Station (BS), and can be based on the quality of service (QoS) requirements of the individual users. Because of high spectral efficiency, as well as robustness against inter-symbol interference, OFDMA was a design choice of recent wireless networking standards, e.g. the IEEE 802.16~\cite{nuyami_wimax_book}, the IEEE 802.20~\cite{ieee80220} and 3GPP Long Term Evolution~\cite{astely_commag_2009}. As opportunistic spectrum use can be implemented efficiently with OFDMA, it seemed natural to connect the strengths of OFDMA with the flexibility of OSA. The first paper that introduced such concept, denoted in the reminder of this paper as Opportunistic Spectrum OFDMA (OS-OFDMA) was~\cite{weiss_commag_2004} (referred therein as Spectrum Pooling) where OFDM subcarriers assigned to individual OSA users (denoted as Secondary Users (SUs)) are deactivated whenever the Primary User (PU) of the radio frequency band reappears. For a recent discussion on the topic of OS-OFDMA please refer to~\cite{mahmoud_wcom_2009}.

So far no theoretical work on the system-level and cross-layer performance of OS-OFDMA networks has been reported. The need for theoretical framework for OS-OFDMA is also motivated by the recent introduction of OSA network standard IEEE 802.22~\cite{ieee80222,cordeiro_book10,stevenson_commag09}. The IEEE 802.22, an extension of the IEEE 802.16 standard, is designed to operate in the vacant TV bands\footnote{Note that the IEEE 802.22 is not the only networking standard proposed that focused on the operation in the TV white spaces. The remaining are recently published ECMA TC48-TG1 standard~\cite{wang_dyspan_2010} focusing on porting local area networks to TV white spaces, and recently started IEEE 802.11af~\cite{ieee80211af}, similar in scope to the aforementioned ECMA activity.}. In the application domain the IEEE 802.22 has already been proposed to bridge remote wireless sensor networks with the command center~\cite{durresi_tranbroad_07} or support Internet connectivity in the rural areas~\cite{liang_wcommag08}. 

Our goal is to develop the analytical framework to analyze the impact of traffic characteristics of OS-OFDMA network subscribers, the activity of the PUs of the radio spectrum, as well as the spectrum sensing algorithm and OFDM subcarrier assignment algorithm on the average throughput of OS-OFDMA network. Our approach is based on a cross-layer Markov chain-based analysis of OS-OFDMA, which allows the investigation of interactions between medium access control and spectrum sensing layer. Since many options of OS-OFDMA subcarrier and subchannel assignment algorithms exist (namely, non-continuos subchannel assignment, as advocated by~\cite{weiss_commag_2004}, and continuos version, as used in the IEEE 802.22~\cite{ieee80222}) it it is important to compare these designs using common analytical framework. In addition, analysis of two-stage spectrum sensing algorithm, proposed by the IEEE 802.22 standard, is challenging due to its complex effect on the medium access control layer and has not been explored in the context of OS-OFDMA communication. Finally, in the context of the IEEE 802.22 analysis could provide estimates of what QoS classes can be supported in OS-OFDMA, given realistic network conditions (such as number and type of primary users, the number and type of QoS classes enabled by OS-OFDMA network and the priority order in channel access for each class of users).

Considering related work, in~\cite{leu_procieee_2009} a general framework of the IEEE 802.16 with OSA capabilities has been proposed with a very simplified networking model, based on Erlang-B formula~\cite[Sec. V-A]{leu_procieee_2009}, where the focus of the paper has been mostly on propagation calculations, including coverage, interference and protection distances. In~\cite{song_chinacom_2008} a simulation platform for the IEEE 802.22-like network, with limited set of ODFMA design options, has been presented. Focusing on the IEEE 802.22, interestingly, while many papers analyzed a certain functionality of the IEEE 802.22 network, like efficient spectrum sensing algorithm design~\cite{chen_globecom_2007,chen_icc_2007}, circuit design for dedicated spectrum sensing~\cite{Park_jssc_09}, MIMO extensions for the IEEE 802.22 physical layer~\cite{kim_casii_2008}, game theoretic analysis of the the IEEE 802.22 networks coexistence~\cite{niyato_ieeewcm_2009,al-zubi_globecom_2009,huang_icics_2009,sengupta_globecom_2008},~\cite{kim_infocom_2010,ko_tvt_2010} (with joint resource allocation), duplexing schemes~\cite{hu_commmag07} (frequency hopping operation),~\cite{liang_wcommag08} (time division duplex design), and mesh establishment~\cite{segupta_iccworkshop_2009}, it is desirable to develop unifying model that captures the IEEE 802.22 intrinsic features such as multiple classes of traffic, two stage spectrum sensing, different types of PUs and their temporal activity, and OFDMA subcarrier allocation process. 

The work closest to the scope of this paper can be found in~\cite{elayoubi_ton08} in which the IEEE 802.16 system was evaluated. Obviously the model developed therein cannot be used directly to evaluate OS-OFDMA system due to the lack of spectrum sensing and PUs activity features. A work similar in our scope (analyzing the system level aspects of subchannel/subcarrier allocation strategies for OS-OFDMA) has been presented in~\cite{luo_twc_2008}. However, no comparison with the the IEEE 802.22 subchannel assignment has been considered. Furthermore, no QoS classes, PU priorities and two-stage spectrum sensing mechanisms were included in the model.

Finally, we need to mention a set of papers analyzing performance of medium access control protocols for OSA networks. Some of the relevant ones include~\cite{Xing_jsac_2006,pawelczak_tvt_2009,Chou_jsac_2007,Park_arxiv_2009}, however none of those works consider OFDMA, usually abstracting underlying physical channel structure.

In this paper, we propose an analytical framework to quantitatively assess the performance of a network based on OS-OFDMA, considering features such as channelization structure, subcarrier allocation, resource assignment to network subscribers and different spectrum sensing methods. In the model we consider different priorities and channel dwell times of SUs and PUs of the spectrum. The developed model allows to calculate capacity (measured in terms of average throughput) for real-time and non-real time QoS classes of OS-OFDMA network.

The rest of the paper is organized as follows. System model describing OS-OFDMA network design options in detail is presented in Section~\ref{sec:system_model}. Analytical model for evaluating throughput of the considered system is presented in Section~\ref{sec:analysis}. Numerical results follow in Section~\ref{sec:numerical_results}. Finally the paper is concluded in Section~\ref{sec:conclusions}.

\section{System Model}
\label{sec:system_model}

\begin{figure}
\centering
\includegraphics[width=\columnwidth]{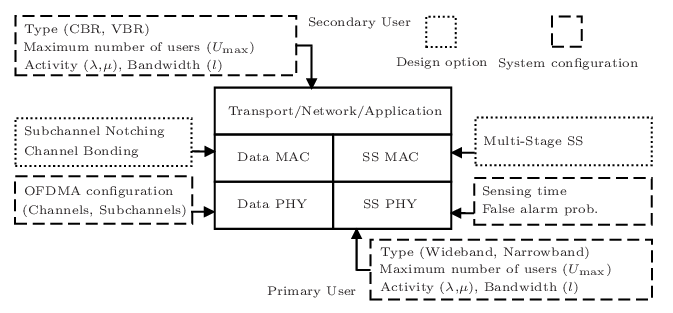}
\caption{OS-OFDMA system model; SS: spectrum sensing.}
\label{fig:SystemModel}
\end{figure}

We consider a centrally controlled network with OS-OFDMA, where a BS manages resources and coordinates spectrum sensing of individual OS-OFDMA network subscribers. Although the proposed model is applicable to both uplink and downlink traffic, for simplicity we assume that only downlink traffic is transmitted. In this paper, we constrain ourselves to a single cell configuration with multiple SUs and multiple PUs, belonging to different user classes. This allows the exclusion of the effect co-channel and adjacent channel interference, as well as co-existence mechanisms in OSA network, on the investigation of the relation between PU type, its activity level and OS-OFDMA design options. On the other hand, we consider transmission errors due to fading and noise on the subchannels. 

A model of the OSA protocol stack under consideration is depicted in Fig.~\ref{fig:SystemModel}. We identify two main modules: (i) data transmission, which is responsible for regular data communication between SUs, and (ii) spectrum sensing, which is responsible for efficient detection of spectrum opportunities for the OSA network; see also~\cite{Park_arxiv_2009} for a similar model. Each component has its unique physical (PHY) and medium access control (MAC) layer. Obviously each layer has its unique design options, for example channel and subchannel management algorithms and multi-stage sensing. Also OSA network can be described by individual parameters such as type of traffic, activity level and bandwidth used. In Section~\ref{sec:system_configuration} we introduce specific OS-OFDMA system configurations considered in this paper. Later in Section~\ref{sec:design options} we introduce the spectrum sensing and MAC design options of interest. We aim to calculate the average throughput obtained at the data MAC layer for all classes of SU traffic, which will be described in detail in Section~\ref{sec:user_types}.

\subsection{OSA System Configuration}
\label{sec:system_configuration}

\subsubsection{Channel Setup and its Relation to OFDMA}
\label{sec:ofdma_channel_config}

The frequency domain consists of $X$ channels, each of which is composed of $Y$ OFDMA subchannels. The total number of subchannels is thus $M=XY$. Each subchannel is further composed of OFDM subcarriers. In this paper, we constrain ourselves to subchannel domain analysis. Furthermore, we assume that subchannel throughput $C$ is on average constant, while its average value depends on the physical layer characteristics such as modulation, error control coding, and MAC layer overhead such as the OFDMA preamble length.

In the time domain, transmission segments are divided in frames of length $t_f$. At the beginning of each frame, SUs of OSA network detect the presence of the PU. We implicitly assume a synchrony between PU and SU activity, as it is a well established and classical assumption in the theoretical analysis, see for example recent publications of~\cite{Park_arxiv_2009},~\cite[Fig. 2]{Liang_twc_2008},~\cite[Sec. III]{pawelczak_tvt_2009},~\cite[Fig. 1]{Hoang_twc_2009},~\cite[Fig. 2]{Papadimitratos_commag_2005}~\cite{gronsund_pimrc_2009,zhang_tvt_2011,zhang_tvt_2009,cheng_icc_2008,zhang_comml_2009,luo_twc_2008}  that follow the same path. Note, however, that the proposed model is extendable to the case where PU slots are offset in time from SU slots. This would require further analysis of PU channel access policies~\cite{Gerihofer_commag_2007,Huang_infocom_2009} which is beyond the scope of this paper. We emphasize, as in~\cite{gambini_twc_2008,Park_arxiv_2009}, that assumption about the synchrony between PU and SU connections allows to calculate throughput upper bounds compared to transmission on a slot-asynchronous interface. Furthermore, we assume that each node in the OSA network observes the same signal emitted by the PU, thus each SU performs all the sensing measurements individually and sends the measurement result to BS on the uplink. Then, BS makes a final decision about the presence of PU on each subchannel.

\subsubsection{PU and SU Types}
\label{sec:user_types}

We consider different types of PUs and SUs. For the PUs, depending on the bandwidth and the activity level, we classify them into: (i) a wideband PU (WPU) having low activity, i.e. with long busy and long idle times, and (ii) a narrowband PU (NPU) having high activity, i.e. with short busy and shorter than WPU idle times. This classification makes the analysis more detailed and realistic. It also makes different scenarios of interest possible to analyze. For example, WPU can represent wireless video links, while NPU can represent wireless microphones, both operating in the TV bands. For SUs, again making our framework general and applicable to multiple scenarios, we consider two types of users: (i) those receiving real-time traffic, denoted as the constant bit rate (CBR) SUs, and (ii) those receiving elastic traffic, denoted as variable bit rate (VBR) SUs, which are included in the the IEEE 802.22 standard. In our OSA network model, different types of SU traffic flows are generated at the upper layers, i.e. application, network and transport, and forwarded down to data PHY, while PU signals are detected at the spectrum sensing module, see Fig.~\ref{fig:SystemModel}.

Furthermore, we assume that a hierarchical structure of users is present, such that the WPU has the highest priority in accessing bandwidth, NPU is the second in access hierarchy, followed by CBR and finally VBR. In other words, if users of different classes could access the same subchannel, the lower priority class user must vacate in order for the higher priority class user to utilize the subchannel. The evacuated CBR switches to the other idle subchannels or drops the connection if there is no idle subchannel available. For VBR, if the PU is detected, the active VBR connection squeezes the bandwidth~\cite[Sec. III-B]{elayoubi_ton08} excluding the channel or subchannel occupied by the PU, and if there is no channel detected as idle, it buffers data until the PU disappears. Note that the behavior of VBR promises to obtain highest possible throughput, as demonstrated in~\cite[Sec. V-B and Fig. 6]{Park_arxiv_2009}, assuming no switching overhead, while CBR does not consider buffering due to the excessive delay that this class might experience while waiting for WPU or NPU to vacate the bandwidth.

Because of the finite channel capacity, the number of users considered in the system is finite, but different for different user types. We assume that at most $U_{w,\max}$, $U_{n,\max}$, $U_{c,\max}$ and $U_{v,\max}$ of WPUs, NPUs, CBR and VBR connections, respectively, can be active at the same time in the considered bandwidth~\cite[Sec. III-B]{elayoubi_ton08}. For the data traffic of SUs and PUs, for analytical tractability, we assume that all users generate new connections according to the negative exponential distribution for the inter-arrival time and burst departure time, which is again a classical assumption in performance analysis studies~\cite{vanmieghem_book_2006}. The analysis can be extended to the general distributions of PU and SU traffic, which is beyond the scope of this paper. Note, however, that the recent measurement campaign~\cite{wellens_phycom_2009} showed that more than 60\% of measured PU activities distributions, including ISM and cellular bands, fitted an exponential distribution. The average inter-arrival and departure time are denoted, respectively, as $1/\lambda_w$ and $1/\mu_w$ for WPU, $1/\lambda_n$ and $1/\mu_n$ for NPU, $1/\lambda_c$ and $1/\mu_c$ for CBR, and  $1/\lambda_v$ and $1/\mu_v$ for VBR. Also we assume that connection of each class except for VBR occupies a fixed number of subchannels. We denote the instantaneous number of subchannels utilized by a connection class as $l_w$ for WPU, $l_n$ for NPU, $l_c$ for CBR and $l_v$ for VBR. Note that the number of subchannels assigned to every connection is fixed and $l_c$, $l_w$, $l_n\in\mathbb{N}$, except for a VBR connection. In that case $l_v\in\mathbb{R}$, which stems from the fact that one data frame consists of a group of OFDMA symbols and the symbols in the OFDMA frame can be assigned to multiple VBR connections. Also, for VBR connections, the burst departure time depends on the number of subchannels used by VBR, thus $1/\mu_v$ is an average departure time when one subchannel is assigned to the VBR connection.

\subsection{Design Options}
\label{sec:design options}

\subsubsection{Spectrum Sensing PHY and MAC Layers}
\label{sec:sensing_phy_mac}

Sensing PHY senses the PU signal and passes the measurement about subchannel availability to the sensing MAC layer for further processing. When a PU is present on the subchannel, it transmits a signal with a certain power and/or unique feature. Thus by detecting the power and/or the feature, the SU can decide whether the PU signal is on the subchannel or not. The main parameters for the spectrum sensing PHY are the probability of detection, the probability of false alarm and the sensing time. There is a trade-off between the sensing time and the resulting probabilities~\cite{peh_tvt_2009}. That is, if a SU takes a long time to sense a subchannel, the time for data transmission may be reduced. However, more idle subchannels can be detected because of high accuracy, which in turn may increase bandwidth utilization. Therefore, sensing time and sensing accuracy are the critical design options for the sensing PHY. 

In the sensing MAC layer, the SUs decide collectively, with the help of BS, on the PU state on the subchannel based on the sensing results of the sensing PHY layer. We denote the detection based on multiple users as collaborative sensing and that based on multiple periods as multi-stage sensing. Since the performance of collaborative sensing is relatively well known, see for example~\cite{Park_arxiv_2009}, in this paper, we focus mostly on multi-stage sensing\footnote{We do not focus on recently proposed spectrum sensing methods for OFDMA networks, like~\cite{song_twc_2010} where quite-active sensing is proposed with non-active users sensing while others actively communicating, since they belong to a single-stage spectrum sensing category.}. For the first results of multi-stage spectrum sensing in network context refer to~\cite{luo_twc_2009,jeon_tvt_2010,Gabran_arxiv_2010} for two-stage sensing multi-channel system, and~\cite{jeon_twc_2008} for two-stage sensing single channel system. 

In this work we limit ourselves to two-stage sensing, noting that our analysis can be directly extended to multi-stage sensing. The procedure works as follows. First, the SU senses the subchannel coarsely at the beginning of every frame, with short sensing time and low sensing accuracy. If the PU is detected, the SU switches to fine sensing mode (immediately, in the same frame), with long sensing time and high sensing accuracy. Depending on the sensing strategy, fine sensing can increase sensing accuracy~\cite{jeon_twc_2008} or frequency resolution~\cite{luo_twc_2009}. Two-stage sensing can be described by different sensing PHY parameters for each stage. We denote the probability of detection as $\delta_a$ and $\delta_{f}$, the probability of false alarm as $\phi_{a}$ and $\phi_{f}$, and the sensing time as $\tau_{a}$ and $\tau_{f}$ for coarse and fine sensing stages, respectively. When setting $\delta_a=1$, $\phi_{a}=1$ and $\tau_{a}=0$ the two-stage sensing model reduces to a single stage sensing model.

In order to evaluate the effect of spectrum sensing on the system throughput, we consider two unique sensing strategies. Firstly, we consider a sensing strategy where the SU senses all channels, including the operating channel, with coarse sensing and when the SU detects the PU on any of the channels, it immediately switches to fine sensing. We name this strategy as general two-stage sensing and denote it as S$_0$.

Secondly, for a specific case when the bandwidth to transmit data is fixed and less than the whole allowed bandwidth, we investigate the following strategy. During coarse sensing the SU senses not the whole bandwidth but only a fixed bandwidth that is currently utilized for data transmission. If the PU is detected on the channel, the SU immediately senses all channels allowed to be utilized for the OSA system with fine sensing and switches to one of the the channels detected as idle. We name this strategy as two-stage active channel sensing and denote it as S$_1$. Since in this strategy, in contrary to S$_0$,  there is no need to always sense all channels, the sensing time is reduced.

\subsubsection{Data PHY and MAC Layer}
\label{sec:data_mac}

Even though the OSA network is aware of subchannels being idle or busy, it should determine how to utilize the subchannels detected as idle for data transmission. There are numerous methods to utilize the idle subchannels in an OSA context, for a good overview we refer to~\cite{Park_arxiv_2009}. However, we selectively study four strategies that are proper for the centrally-controlled OS-OFDMA-based network. Those four strategies can be classified into two groups based on the purpose of the strategies.

Firstly, we need to determine how much bandwidth is utilized for data transmission from the channels detected as idle. One strategy is to utilize all channels detected as idle from the allowed bandwidth. This strategy may maximize channel utilization at the cost of the wideband RF and signal processing. We name this strategy as variable channel/subchannel bonding and denote it as B$_1$. On the other hand, another strategy is to transmit data through only one channel, even though there may exist more channels detected as idle. Because the SU operates on the bandwidth of only one channel, the cost for the RF and signal processing is low. However, it is inefficient because some available bandwidth may not be utilized. Since it is one of the operating modes of the IEEE 802.22, also advocated by Federal Communications Commission~\cite{fcc_docket_2008}, we also include it in our study. We name this strategy fixed channel selection and denote it symbolically as B$_0$.

Secondly, we also need a strategy to avoid utilizing the subchannels on which the PU is detected. An efficient strategy is to notch out the subchannel detected as busy and utilize all other available subchannels for OS-OFDMA. We name this strategy subchannel notching and denote it as N$_1$. Note that we assume for simplicity that it is possible to notch out subchannels and transmit on the adjacent subchannels without causing interference, which is a common assumption in system level analysis, e.g.~\cite{zhang_tvt_2011,zhang_tvt_2009,luo_twc_2008,ngo_tvt_2010}\footnote{Please refer to a recent studies on that topic that prove the feasibility of such approach. For example, in~\cite{qu_tvt_2010} a sidelobe suppression with guard band equal to only one OFDM subcarrier interval was shown. In~\cite{yuan_tvt_2010} a OFDM subcarrier notching was proposed with only 4\% of the available spectrum wastage. Finally a practical implementation of OFDM subcarrier suppression with perfect channel utilization at the cost of throughput reduction was demonstrated in~\cite{sutton_dyspan_2010}.}. On the opposite side, a conservative strategy is to exclude (block) all subchannels within the operating channel from accessing, even though only one subchannel is utilized by the PU, which is suggested for the IEEE 802.22. We name this strategy channel blocking and denote it symbolically as N$_0$.

Please note that channel switching delays is in the order of tens to hundreds of microseconds~\cite{sharma_jsac2004,raniwala_infocom_2005} which is a fraction of the channel sensing time. Dependent on protocol design, an additional OFDMA frame might be needed to communicate the decision about channel switch between base station and terminals. For simplicity channel switching delay is neglected in the analysis.

\subsubsection{Design Options of Interest}
\label{sec:design_options_interest}

Because we have three groups of binary choices, i.e. S$_x$, N$_x$, and B$_x$, where $x\in\{0,1\}$, there can be eight possible combinations of design options. However, not all options are feasible. First, we do not consider the combination of subchannel notching (N$_1$) and fixed channel selection (B$_0$), since it is a special case of N$_1$B$_1$ configuration with a single channel. Also, for two-stage active channel sensing (S$_1$), we only consider channel blocking (N$_0$) and fixed channel selection (B$_0$), because two-stage active channel sensing (S$_1$) is applicable to fixed bandwidth utilization case only. This leaves four combinations of the design options which are summarized in Table~\ref{tab:options}. Note, that all options except S$_0$N$_1$B$_1$, are considered by the IEEE 802.22 standard.

\begin{table}
\centering
\caption{Summary of Design Options of Considered in The Paper}
\begin{tabular}{c|c|c|c}
\hline
Symbol & Two-Stage Sensing & Notching/Blocking & Bonding/Separation \\
\hline\hline
S$_0$N$_1$B$_1$ & General & Subchannel Notching & Subchannel Bonding\\
S$_0$N$_0$B$_1$ & General & Channel Blocking & Channel Bonding\\
S$_0$N$_0$B$_0$ & General & Channel Blocking & Fixed Channel Selection\\
S$_1$N$_0$B$_0$ & Active Channel & Channel Blocking & Fixed Channel Selection\\
\hline
\end{tabular}
\label{tab:options}
\end{table}

\section{Proposed Analytical Model}
\label{sec:analysis}

In this section we will describe the calculation of average throughput obtained using each of the considered OS-OFDMA designs. The analysis is based on a probabilistic framework utilizing Markov chain. We start with S$_0$N$_1$B$_1$ option, which serves as a foundation to analyze the remaining three OS-OFDMA designs.

\subsection{Case S$_0$N$_1$B$_1$ (General Sensing, Subchannel Notching, Subchannel Bonding)}
\label{sec:s0n1b1}

Generally, in OFDMA-based wireless networks throughput depends on how many subchannels are utilized in the idle spectrum by each SU connection type~\cite[Sec. III-B]{elayoubi_ton08} (in case of our model: by every CBR and VBR connection). Let $\Pr_{1}(m_c,m_v,M_a)$ be the probability that $m_c$ and $m_v$ subchannels are utilized by CBR and VBR connections, respectively\footnote{In the paper we follow the convention that each newly introduced probability will be uniquely identified by a number and introduced with all argument variables, while its later callouts will be referred as $\Pr_{x}(\cdot)$.} when $M_a$ subchannels are detected as idle. In addition, let $\eta(M_a)$ be the average ratio of data transmission time to total frame length when $M_a$ subchannels are detected as idle. Then, the total system throughput, $H$, can be calculated as
\begin{equation}
H\triangleq C \sum_{M_a=0}^M\eta(M_a)\!\!\sum_{m_c,m_v \in \{0,\cdots,M\}}(m_c + m_v){\Pr}_{1}(m_c,m_v,M_a).
\label{eq:H}
\end{equation}
In addition, the throughput of CBR connection, $H_c$, and VBR connection, $H_v$, is computed using (\ref{eq:H}) by replacing $m_c+m_v$ with $m_c$ for $H_c$ and $m_v$ for $H_v$. In the subsequent sections we will describe a method to derive $\Pr_{1}(\cdot)$ and $\eta(\cdot)$. In this method we will hierarchically decompose $\Pr_{1}(\cdot)$  and $\eta(\cdot)$ into a set of conditional probabilities. Each probability will describe a particular relation between spectrum sensing outcome, state of CBR and VBR connection and the PU activity.

\subsubsection{Derivation of Probability of Number of CBR and VBR Connections and Available Subchannels, $\Pr_{1}(\cdot)$}

Values of $m_c$ and $m_v$ are easily determined if we know the number of CBR connections connected with the BS, $U_c$, the number of VBR connections connected with the BS, $U_v$, and the number of subchannels detected as idle, $M_a$. Thus
\begin{align}
{\Pr}_{1}(\cdot)\triangleq\sum_{U_c=0}^{U_{c,\max}} \sum_{U_v=0}^{U_{v,\max}}{\Pr}_{2}(m_c,m_v|U_c,U_v,M_a){\Pr}_3(U_c,U_v,M_a),
\label{eq:P1}
\end{align}
where $\Pr_{2}(\cdot)$ denotes a set of allowed subchannel configurations occupied by CBR and VBR connections and $\Pr_{3}(\cdot)$ denotes the probability of active $U_c$ CBR connections, $U_v$ VBR connections and $M_a$ idle subchannels.

\paragraph{Derivation of Allowed CBR and VBR Subchannel Configurations, $\Pr_{2}(\cdot)$}

Since the total number of subchannels used by all CBR users, $U_c l_c$, cannot be greater than the number of the available subchannels, $M_a$, because if there is no subchannel available the CBR connection will be blocked, the only valid case is $U_c l_c\leq M_a$. Furthermore, a CBR connection has a higher priority than a VBR connection, all CBR connections can transmit data through all $m_c=U_c l_c$ subchannels. Then the remaining subchannels, i.e. $m_v=M_a-U_c l_c$, are used by VBR connections. On the other hand, if there are no VBR connections in the system then $m_v=0$. Therefore, defining $\mathbb{U}(x)$, where $\mathbb{U}(x\leq0)=0$ and $\mathbb{U}(x>0)=1$ we have
\begin{equation}
{\Pr}_{2}(\cdot)\triangleq
\begin{cases}
1,& \begin{split}U_c l_c\leq M_a, m_c=U_cl_c,m_v=(M_a-U_cl_c)\mathbb{U}(U_v);\end{split}\\
0,& \begin{split}\text{otherwise}\end{split}.
\end{cases}
\end{equation}

\paragraph{Derivation of Probability of Active CBR and VBR Connections and Idle Subchannels, $\Pr_{3}(\cdot)$}

Since our model considers arrival process of SU connections, which departure process is affected by PU temporal activity, to calculate $\Pr_3(\cdot)$ in (\ref{eq:P1}), we need a tool to evaluate steady state probability of given number CBR and VBR connections, as well as the number of idle subchannels. To do this we use a widely used method based on the composition of a Markov chain~\cite[Ch. 11]{vanmieghem_book_2006}. 

We introduce a Markov chain state $\{U_c,U_v,M_a\}$. Furthermore, we introduce the state transition probability which describes the change in $\{U_c,U_v,M_a\}$ between time $t-1$ and $t$, denoted as $\Pr_{4}\left(\cdot\right)$. Then we can compute $\Pr_3(\cdot)$ by solving the Markov chain, given $\sum_{\mathcal{M}}\Pr_3(\cdot)=1$ and $\Pr_3(\cdot)=\sum_{\mathcal{M}^{(t-1)}}\Pr_3(\cdot)\Pr_4(\cdot)$, where $\mathcal{M}$ is a set of all possible states $\{U_c,U_v,M_a\}$ and $\mathcal{M}^{(x)}$ is a set of the states at time $x$. Based on the conditional probability property and independency of the variables, we decompose $\Pr_4(\cdot)$ as
\begin{align}
{\Pr}_{4}\left(\cdot\right)\triangleq{\Pr}_{5}\left(U_{c}^{(t)},U_{v}^{(t)}|U_{c}^{(t-1)},U_{v}^{(t-1)},M_{a}^{(t)},M_{a}^{(t-1)}\right){\Pr}_{6}\left(M_{a}^{(t)}|M_{a}^{(t-1)}\right),
\label{eq:P4}
\end{align}
where $\Pr_{5}(\cdot)$, denotes the probability of change in number of CBR and VBR connections count, and $\Pr_{6}(\cdot)$, denotes the probability of change in number of subchannels detected as idle. We derive those expressions below.

\paragraph{Derivation of Probability of Change in CBR and VBR Connection Count, $\Pr_{5}(\cdot)$}

We can decompose $\Pr_{5}(\cdot)$ into probabilities denoting a change in the number of connections separately for CBR and VBR. Accordingly, we define ${\Pr}_{7}\left(U_{c}^{(t)}|U_{c}^{(t-1)},M_{a}^{(t)},M_{a}^{(t-1)}\right)$ as the conditional probability of the number of the CBRs at time $t$ for the given number of subchannels detected as idle, and ${\Pr}_{8}\left(U_{v}^{(t)}|U_{v}^{(t-1)},U_{c}^{(t)},U_{c}^{(t-1)},M_{a}^{(t)},M_{a}^{(t-1)}\right)$ as the conditional probability of the number of VBR connections at time $t$ for the given number of CBR connections and the available subchannels. Note that in $\Pr_{7}(\cdot)$, since we assume that the CBR has higher priority than the VBR and the VBR connections utilize the remaining subchannels after subchannel assignment for all CBR connections, there is no dependency on $U_{v}^{(t-1)}$. Furthermore, note that in $\Pr_{7}(\cdot)$ and $\Pr_{8}(\cdot)$ change in the number of CBR and VBR connections, respectively, depends on the number of subchannels detected as idle at time $t-1$ and $t$. Then we have
\begin{align}
{\Pr}_{5}\left(\cdot\right)\triangleq&\;{\Pr}_{7}\left(U_{c}^{(t)}|U_{c}^{(t-1)},M_{a}^{(t)},M_{a}^{(t-1)}\right){\Pr}_{8}\left(U_{v}^{(t)}|U_{v}^{(t-1)},U_{c}^{(t)},U_{c}^{(t-1)},M_{a}^{(t)},M_{a}^{(t-1)}\right).
\label{eq:P5}
\end{align}
We proceed with describing the process of deriving the expressions for $\Pr_{7}(\cdot)$ and $\Pr_{8}(\cdot)$.

\paragraph{Derivation of Probability of Change in CBR Connection Count, $\Pr_{7}(\cdot)$}

First, we consider valid conditions for $U_{c}^{(t)}$, $U_{c}^{(t-1)}$, $M_{a}^{(t)}$, and $M_{a}^{(t-1)}$ for $\Pr_7(\cdot)$ involving all possible number of $i$ arriving and $j$ departing connections. We denote the number of users being able to utilize all available subchannels as $U_{a}^{(t)}=\lfloor M_{a}^{(t)}/l_c \rfloor$ at time $t$ and $U_{a}^{(t-1)}=\lfloor M_{a}^{(t-1)}/l_c \rfloor$ at time $t-1$. Then, because $U_{a}^{(t)}$, $U_{a}^{(t-1)}$ denote the maximum number of users $U_{c}^{(t)}\leq U_{a}^{(t)}$ and $U_{c}^{(t-1)}\leq U_{a}^{(t-1)}$. Furthermore, because the possible sets of $i$ and $j$ are different for the cases $U_{c}^{(t)}< U_{a}^{(t)}$, $U_{c}^{(t)} = U_{a}^{(t)}>0$, and $U_{c}^{(t)} = U_{a}^{(t)}=0$, we consider them separately.

The first case, $U_{c}^{(t)}<U_{a}^{(t)}$, represents the situation when the number of subchannels detected as idle is more than the number of all subchannels that will be utilized by CBR connections before spectrum sensing. In other words, no CBR connection is blocked due to the PU appearance. Because the number of CBR connections is $U_{c}^{(t-1)}$ at time $t-1$ and $U_{c}^{(t)}$ at time $t$, the change in the number of CBR connections is $i-j=U_{c}^{(t)}-U_{c}^{(t-1)}$. In addition, because there are $U_{c}^{(t-1)}$ active connections at time $t-1$, more than $U_{c}^{(t-1)}$ connections cannot be released, i.e. $j\le U_{c}^{(t-1)}$. Therefore $\{i,j\}\in \mathcal{K}_{c,a}\triangleq\left\{i,j|i-j=U_{c}^{(t)}-U_{c}^{(t-1)}, j\le U_{c}^{(t-1)}\right\}$.

The second case, $U_{c}^{(t)}=U_{a}^{(t)}>0$, denotes the situation when CBR connections may be blocked due to the PU arrival. Then, before spectrum sensing, the total number of connections including newly generated connections is $U_{c}^{(t-1)}+i-j$. However, after spectrum sensing, the connections that utilize subchannels detected as busy are blocked and the remaining connections, $U_{c}^{(t)}$, utilize all available subchannels, $M_{a}^{(t)}$. Thus, the number of CBR connections before spectrum sensing, $U_{c}^{(t-1)}+i-j$, can be greater than or equal to the number of CBR connections after spectrum sensing, $U_{c}^{(t)}$, but should be less than or equal to $U_{c,\max}$, i.e. $U_{c}^{(t)}\leq U_{c}^{(t-1)}+i-j \leq U_{c,\max}$. Therefore $\{i,j\}\in \mathcal{K}_{c,b}\triangleq\left\{i,j|U_{c}^{(t)}-U_{c}^{(t-1)}\leq i-j \leq U_{c,\max}-U_{c}^{(t-1)}, j\le U_{c}^{(t-1)}\right\}$.

The third and final case is when $U_a^{(t)}=U_c^{(t)}=0$. In this situation, because there is no subchannels available for CBR, the number of CBR connections should also be zero. Consequently, regardless of $i$ and $j$, the conditional probability $\Pr_{7}(\cdot)$ under this condition is always one. 

Now, we introduce two supporting functions, $G_x(i|U_x,\lambda_x)$ and $T_x(j|U_x,\mu_x)$, which will be used to derive $\Pr_{7}(\cdot)$, where $G_x(i|U_x,\lambda_x)$ is the probability that $i$ connections are newly generated from $U_x$ available users with arrival rate $\lambda_x$, and $T_x(j|U_x,\mu_x)$ is the probability that $j$ connections are released, each with departure rate $\mu_x$. $G_x(i|U_x,\lambda_x)$ and $T_x(j|U_x,\mu_x)$ are derived in Appendix \ref{app:TG}. Our approach to derive $\Pr_7(\cdot)$ is to calculate all possible sets for $i$ and $j$ and applying them to $G_x(i|U_x,\lambda_x)$ and $T_x(j|U_x,\mu_x)$. In result, $\Pr_7(\cdot)$ is derived as
\begin{equation}
{\Pr}_{7}(\cdot)\triangleq
\begin{cases}
\sum_{\{i,j\} \in \mathcal{K}_{c,a}} T_{c}(j|U_{c}^{(t)},\mu_c) G_{c}\left(i|U_{c}^{(t-1)},\lambda_c\right), & \begin{split}&U_{c}^{(t-1)}\leq U_{a}^{(t-1)},\\&U_{c}^{(t)}<U_{a}^{(t)};\end{split}\\
\sum_{\{i,j\} \in \mathcal{K}_{c,b}}  T_{c}(j|U_{c}^{(t)},\mu_c) G_{c}\left(i|U_{c}^{(t-1)}+i-j,\lambda_c\right), & \begin{split}U_{c}^{(t-1)}\leq U_{a}^{(t-1)},\\U_{c}^{(t)}=U_{a}^{(t)}>0;\end{split}\\
1, & \begin{split}U_{c}^{(t-1)}\leq U_{a}^{(t-1)},\\U_a^{(t)}=U_c^{(t)}=0.\end{split}
\end{cases}
\label{eq:P7}
\end{equation}

\paragraph{Derivation of Probability of Change in VBR Connection Count, $\Pr_{8}(\cdot)$}

In our model we do not consider the case that the VBR connection is blocked by the PU because VBR connections are assumed to be buffered instead of blocked when there is no available subchannel. Thus, assuming that all VBR connections share the same portion of the idle bandwidth, we calculate the number of subchannels assigned to one VBR connection, $l_v$, as
\begin{equation}
l_{v}=
\begin{cases}
\frac{M_{a}^{(t-1)}-U_{c}^{(t-1)}}{U_{v}^{(t-1)}},&U_{v}^{(t-1)} > 0,\\
0,&U_{v}^{(t-1)}=0.
\end{cases}
\label{eq:lv}
\end{equation}
In turn, $\Pr_{8}(\cdot)$ from (\ref{eq:P5}) is defined similarly to (\ref{eq:P7}) as
\begin{equation}
{\Pr}_{8}\left(\cdot\right)\triangleq\sum_{\{i,j\} \in \mathcal{K}_v} T_v(j|U_{v}^{(t)},l_{v}\mu_v) G_v\left(i|U_{v}^{(t-1)},\lambda_v\right),
\label{eq:P8}
\end{equation}
where $\mathcal{K}_v\triangleq \left\{i,j|i-j=U_{v}^{(t)}-U_{v}^{(t-1)}, j\le U_{v}^{(t-1)}\right\}$.

\paragraph{Derivation of Probability of Change of Subchannel Count Detected as Idle, $\Pr_{6}(\cdot)$}

Proceeding to derive $\Pr_{6}(\cdot)$ in (\ref{eq:P4}), it can be decomposed as
\begin{equation}
{\Pr}_{6}\left(\cdot\right)\triangleq \frac{{\Pr}_{9}\left(M_{a}^{(t)},M_{a}^{(t-1)}\right)}{{\Pr}_{10}\left(M_{a}^{(t-1)}\right)},
\label{eq:P6}
\end{equation}
where $\Pr_{9}(\cdot)$ denotes the probability that $M_a$ subchannels were detected as idle at time $t-1$ and $t$, and $\Pr_{10}(\cdot)$ denotes the probability that $M_a$ subchannels were detected as idle at time $t-1$. We will explain the derivation of $\Pr_{9}(\cdot)$ and $\Pr_{10}(\cdot)$ in subsequent sections.

\paragraph{Derivation of Probability of $M_a$ Number of Subchannels Detected as Idle at Time $t-1$ and $t$, $\Pr_{9}(\cdot)$}

The idea behind derivation of (\ref{eq:P9}) is that number of detected subchannels depends on what sensing stage was utilized at time slots $t-1$ and $t$ and how many NPUs and WPUs were present at these time slots. It can be defined as
\begin{align}
{\Pr}_{9}\left(\cdot\right)\triangleq\sum_{\substack{\forall U_{w}^{(t)}, U_{n}^{(t)}, S^{(t)},\\U_{w}^{(t-1)},U_{n}^{(t-1)}, S^{(t-1)}}} &{\Pr}_{11}\left(M_{a}^{(t)},M_{a}^{(t-1)}|U_{w}^{(t)}, U_{w}^{(t-1)}, U_{n}^{(t)}, U_{n}^{(t-1)}, S^{(t)}, S^{(t-1)}\right)\nonumber\\
&\quad \times {\Pr}_{12}\left(U_{w}^{(t)}, U_{w}^{(t-1)}, U_{n}^{(t)}, U_{n}^{(t-1)}, S^{(t)}, S^{(t-1)}\right),
\label{eq:P9}
\end{align}
where $S^{(t)}$ and $S^{(t-1)}$ are the sensing stages at times $t$ and $t-1$, respectively. Furthermore, ${\Pr}_{11}\left(\cdot\right)$ denotes the probability of the number of subchannels detected as idle given the number of WPUs and NPUs and the sensing stage at time $t-1$ and $t$, and ${\Pr}_{12}\left(U_{w}^{(t)}, U_{w}^{(t-1)}, U_{n}^{(t)}, U_{n}^{(t-1)}, S^{(t)}, S^{(t-1)}\right)$ denotes the joint probability of the numbers of WPUs and NPUs and the sensing stage at time $t-1$ and $t$. We can further decompose $\Pr_{11}(\cdot)$ and $\Pr_{12}(\cdot)$ into subsequent probabilities. 

\paragraph{Derivation of Probability of Number of Subchannels Detected as Idle Given the Number of WPUs and NPUs and the Sensing Stage at Time $t-1$ and $t$, $\Pr_{11}(\cdot)$}

Probability $\Pr_{11}(\cdot)$ in (\ref{eq:P9}) can be decomposed into products of probabilities describing available number of subchannels detected as idle at time $t$ and time $t-1$, because the number of subchannels detected as idle and the state of the sensing stage at time slot $t$ is independent from time $t-1$. That is
\begin{align}
{\Pr}_{11}\left(\cdot\right)\triangleq{\Pr}_{13}\left(M_{a}^{(t-1)}| U_{w}^{(t-1)}, U_{n}^{(t-1)}, S^{(t-1)}\right) {\Pr}_{13}\left(M_{a}^{(t)}| U_{w}^{(t)}, U_{n}^{(t)},S^{(t)}\right),
\label{eq:P11}
\end{align}
where $\Pr_{13}(\cdot)$ is derived in Appendix~\ref{sec:p13s0n1b1}.

\paragraph{Derivation of Probability of the Numbers of WPUs and NPUs and the Sensing Stage at Time $t-1$ and $t$, $\Pr_{12}(\cdot)$}

Probability ${\Pr}_{12}\left(\cdot\right)$ in (\ref{eq:P9}) can be decomposed into conditional probabilities as follows
\begin{align}
{\Pr}_{12}(\cdot)\triangleq&\;{\Pr}_{15}\left(S^{(t)} | U_{w}^{(t)}, U_{n}^{(t)}\right) {\Pr}_{15}\left( S^{(t-1)}| U_{w}^{(t-1)}, U_{n}^{(t-1)}\right)\nonumber\\&\times{\Pr}_{16}\left(U_{w}^{(t)}, U_{n}^{(t)}| U_{w}^{(t-1)}, U_{n}^{(t-1)}\right){\Pr}_{17}\left(U_{w}^{(t-1)}, U_{n}^{(t-1)}\right),
\label{eq:P12}
\end{align}
where $\Pr_{15}(\cdot)$ is the probability of being in a certain sensing stage given the number of NPUs and WPUs, which is derived in Appendix~\ref{sec:p15s0n1b1}, and ${\Pr}_{16}(\cdot)$ and $\Pr_{17}(\cdot)$ are the state transition probability and the steady state probability for the state $\{U_{w}, U_{n}\}$, respectively. As in the case of $\Pr_{3}(\cdot)$ denoting probability of active number of CBR and VBR connections, since the state of NPU and WPU changes randomly and independently from time slot to time slot, we can apply the same method of Markov chain construction to derive the probability of being in any of the $\{U_{w}, U_{n}\}$ states. We will describe their derivation below.

\paragraph{Derivation of Steady State Transition Probability for State $\{U_w,U_n\}$, $\Pr_{17}(\cdot)$}

Probability $\Pr_{17}(\cdot)$ in (\ref{eq:P12}) is computed by solving a Markov chain, such that $\sum_{\mathcal{U}}\Pr_{17}(\cdot)=1$ and $\Pr_{17}(\cdot)=\sum_{\mathcal{U}^{(t-1)}}\Pr_{17}(\cdot)\Pr_{16}(\cdot)$, where $\mathcal{U}$ is the set of all possible values of $U_{w}$ and $U_{n}$, and $\mathcal{U}^{(t-1)}$ is the set of the same parameters at time $t-1$. Based on the assumption that the WPU has a higher priority of channel access than the NPU, the state transition probability $\Pr_{16}(\cdot)$ is derived as 
\begin{align}
{\Pr}_{16}\left(\cdot\right)\triangleq\;&{\Pr}_{18}\left(U_{w}^{(t)}| U_{w}^{(t-1)}\right){\Pr}_{19}\left(U_{n}^{(t)}| U_{n}^{(t-1)}, U_{w}^{(t)}, U_{w}^{(t-1)}\right),
\end{align}
where $\Pr_{18}(\cdot)$ denotes the probability of change in number of WPUs, $U_w$, between time slot $t-1$ and $t$, and $\Pr_{19}(\cdot)$ denotes the probability of change in number of NPUs, $U_n$, between time slot $t-1$ and $t$. We describe their derivation below.

\paragraph{Derivation of Probability of Change in the Number of WPUs and NPUs, $\Pr_{18}(\cdot)$ and $\Pr_{19}(\cdot)$}

From the introduced model, the inter-arrival time and departure time follows the negative exponential distribution. Therefore we can use equations (\ref{eq:G}) and (\ref{eq:T}), derived in Appendix~\ref{app:TG}, in the similar way as in the derivation of (\ref{eq:P7}) and (\ref{eq:P8}). By denoting the available subchannels for NPU as $U_{e2}=\left \lfloor (M- U_{w}^{(t)}l_w)/l_n \right\rfloor$ at time $t$ and $U_{e1}=\left\lfloor (M- U_{w}^{(t-1)}l_w)/l_n \right\rfloor$ at time $t-1$, we can derive $\Pr_{18}(\cdot)$ as
\begin{equation}
{\Pr}_{18}\left(\cdot\right)\triangleq\sum_{\{i,j\} \in \mathcal{K}_w} T_w\left(j|U_{w}^{(t-1)},\mu_w\right)G_w\left(i|U_{w}^{(t)},\lambda_w\right),
\label{eq:P18}
\end{equation}
and $\Pr_{19}(\cdot)$ as
\begin{equation}
{\Pr}_{19}\left(\cdot\right)\triangleq\\
\begin{cases}
\displaystyle \sum_{\{i,j\} \in \mathcal{K}_{n,a}} T_n(j|U_{n}^{(t)},\mu_n) G_n\left(i|U_{n}^{(t-1)},\lambda_n\right), &\!\!\!\!\begin{split}U_{n}^{(t-1)}\leq U_{e1},U_{n}^{(t)}<U_{e2};\end{split}\\
\displaystyle \sum_{\{i,j\} \in \mathcal{K}_{n,b}} T_n(j|U_{n}^{(t)},\mu_n) G_n\left(i|U_{n}^{(t-1)}+i-j,\lambda_n\right), &\!\!\!\!\begin{split}U_{n}^{(t-1)}\leq U_{e1},U_{n}^{(t)}=U_{e2},\end{split}
\end{cases}
\label{eq:P19}
\end{equation}
where $\mathcal{K}_{w}\triangleq\left\{i,j|i-j= U_{w}^{(t)}- U_{w}^{(t-1)}, j\le  U_{w}^{(t-1)}\right\}$, $\mathcal{K}_{n,a}\triangleq\left\{i,j|i-j= U_{n}^{(t)}- U_{n}^{(t-1)}\right.,$\\ $\left. j\le  U_{n}^{(t-1)}\right\}$ and $K_{n,b}\triangleq\left\{i,j| U_{n}^{(t)}- U_{n}^{(t-1)} \leq i-j \leq U_{n,\max}- U_{n}^{(t-1)},j\le  U_{n}^{(t-1)}\right\}$.

\paragraph{Derivation of Probability of $M_a$ Subchannels were Detected as Idle at Time $t-1$, $\Pr_{10}(\cdot)$}

Finally, $\Pr_{10}(\cdot)$ in (\ref{eq:P6}) is calculated as
\begin{align}
{\Pr}_{10}\left(\cdot\right)\triangleq\sum_{\forall U_{w}^{(t-1)}, U_{n}^{(t-1)}, S^{(t-1)}}&{\Pr}_{13}(M_{a}^{(t-1)}|U_{w}^{(t-1)},U_{n}^{(t-1)},S^{(t-1)})\nonumber\\&\times{\Pr}_{15}(S^{(t-1)}|U_{w}^{(t-1)},U_{n}^{(t-1)}){\Pr}_{17}( U_{w}^{(t-1)}, U_{n}^{(t-1)}).
\label{eq:P10}
\end{align}

\subsubsection{Derivation of Sensing Overhead, $\eta(\cdot)$}

The value of $\eta(\cdot)$ depends on the sensing stage since a longer sensing time for one stage can reduce $\eta(\cdot)$, while a shorter sensing time for another stage can increase $\eta(\cdot)$. We recall a variable $S$, introduced in Appendix~\ref{sec:p13s0n1b1} and Appendix~\ref{sec:p15s0n1b1}, which indicates the sensing stage, such that $S=0$ denotes the case when the OSA network performs only coarse sensing without switching to fine sensing, and $S>0$ denotes the case when the OSA network performs coarse sensing and switches to fine sensing immediately. Specifically, $S=1$ denotes the case when the OSA network detects the idle subchannel and $S=2$ denotes the case that no idle subchannel is detected so that the network waits until the next sensing period without transmitting data. 

Defining $\Pr_0(S,M_a)$ as the joint probability that the current sensing stage equals to $S$ and the number of subchannels detected as idle is $M_a$, we can compute $\eta$ as
\begin{align}
\eta(M_a)\triangleq\:&{\Pr}_0(0,M_a)\frac{t_f\!-\!\tau_{a}}{t_f}\!+\!{\Pr}_0(1,M_a)\frac{t_f\!-\!\tau_{a}\!+\!\tau_f}{t_f}.
\label{eq:eta}
\end{align}
Note that there is no ${\Pr}_0(2,M_a)$ in (\ref{eq:eta}) because for $S=2$ no data can be transmitted, so the ratio of the data transmission time to the total frame length is zero. Finally, we can derive ${\Pr}_0(\cdot)$ in (\ref{eq:eta}) as (\ref{eq:P10}) removing $S^{(t-1)}$ from the lowest bound of summation.

\subsection{Case S$_0$N$_0$B$_1$ (General Sensing, Channel Blocking, Channel Bonding)}
\label{sec:s0n0b1}

There are two major changes for the analysis of S$_0$N$_0$B$_1$ case in comparison to S$_0$N$_1$B$_1$. First, the number of subchannels detected as idle should be the integer multiples of $Y$ because in this case the smallest quantity of idle bandwidth is one channel. Second, the number of available subchannels is determined not only by the number of the NPUs but also by the position of the NPUs in the spectrum. For example, if two NPUs appear on different subchannels in the same channel, the SU cannot utilize that channel. If NPUs occupy subchannels located in two different channels, those two channels cannot be used by SUs. Considering those changes, we need to modify probabilities related to the number of subchannels detected as idle and the sensing stage used, that is ${\Pr}_{11}(\cdot)$ in (\ref{eq:P11}), ${\Pr}_{13}(\cdot)$ in (\ref{eq:P13}) and ${\Pr}_{15}(\cdot)$ in (\ref{eq:P15}). Note that the other probabilities remain the same as derived in Section~\ref{sec:s0n1b1}. In the process of modification of the above expressions we assume $\delta_f \approxeq \delta_a \approxeq 1$ to reduce the complexity of calculations. In general, the OS-OFDMA system needs to keep a high detection probability to protect the PUs, which makes this approximation reasonable. Note, however, that we will still consider the effect of false alarms. Thus, even if there is no PU on the spectrum, the SU may falsely detect an idle subchannel as busy. Moreover, without loss of generality, we assume that a WPU occupies one channel, i.e., $l_w=Y$.

To modify ${\Pr}_{11}(\cdot)$ from (\ref{eq:P11}) we observe that the change of the number of PUs between time $t$ and $t-1$ may affect the change of the position of the PUs and, as a result, can make an impact on the number of channels detected as idle at time $t$. Thus, introducing the conditional probability of the number of idle subchannels as ${\Pr}_{23}\left(M_{a}^{(t)}| U_{w}^{(t)}, U_{n}^{(t)} ,S^{(t)},M_{a}^{(t-1)},U_{w}^{(t-1)}, U_{n}^{(t-1)} ,S^{(t-1)}\right)$, ${\Pr}_{11}(\cdot)$ in (\ref{eq:P11}) is newly defined as
\begin{align}
{\Pr}_{11}\left(\cdot\right)\triangleq&\;
{\Pr}_{13}\left(M_{a}^{(t-1)}| U_{w}^{(t-1)}, U_{n}^{(t-1)}, S^{(t-1)}\right)\nonumber\\
&\quad\times{\Pr}_{23}\left(M_{a}^{t}| M_{a}^{(t-1)}, U_{w}^{(t)},U_{w}^{(t-1)},U_{n}^{(t)}, U_{n}^{(t-1)}, S^{(t)}, S^{(t-1)}\right),
\label{eq:P11a}
\end{align}
where $\Pr_{23}(\cdot)$ is derived in Appendix~\ref{sec:P23S0N0B1}. Finally, $\Pr_{13}(\cdot)$ and ${\Pr}_{15}(\cdot)$ are modified in Appendix~\ref{sec:P13S0N0B1} and Appendix~\ref{sec:P15S0N0B1}, respectively.

\subsection{Case S$_0$N$_0$B$_0$ (General Sensing, Channel Blocking, Fixed Channel Selection)}
\label{sec:s0n0b0}

The analysis in this section is based on the derivation from Section~\ref{sec:s0n0b1} since this option still considers subchannel non-notching. The major change here is that we perform analysis for data transmission on one channel instead of all $X$ channels. This change affects the valid condition for ${\Pr}_{7}(\cdot)$ in (\ref{eq:P7}) and $l_v$ in (\ref{eq:lv}). Because the maximum number of available subchannels is limited to $Y$, $U_a^{(t)}$ and $U_a^{(t-1)}$ becomes $\left \lfloor \min \left(Y/l_n,M_a^{(t)}/l_n\right) \right\rfloor$ and $\left\lfloor \min \left(Y/l_n,M_a^{(t-1)}/l_n\right) \right\rfloor$, respectively. Moreover, considering that the maximum number of channels utilized by VBR is also limited to one channel, then (\ref{eq:lv}) should be modified as
\begin{equation}
l_{v}=
\begin{cases}
\frac{\min \left(Y,M_{a}^{(t-1)}\right)-U_{c}^{(t-1)}}{U_{v}^{(t-1)}},&U_{v}^{(t-1)} > 0,\\
0,&U_{v}^{(t-1)}=0.
\end{cases}
\label{eq:lva}
\end{equation}

Also, we need to modify ${\Pr}_{6}(\cdot)$ in (\ref{eq:P6}) (probability of change of subchannel count detected as idle) considering the limitations of the available subchannels. Even if a SU detects more than one idle channel, the SU will utilize only one channel. In terms of the definition, we have to sum all probabilities that the number of subchannels detected as idle is greater than or equal to $Y$, in order to compute the probability that one channel, i.e. $Y$ subchannels, are detected as idle. Thus,
\begin{equation}
{\Pr}_{6}(\cdot)\triangleq
\begin{cases}
\frac{{\Pr}_{9}(0,0)}{{\Pr}_{10}(0)},&M_{a}^{(t)}=0,M_{a}^{(t-1)}=0,\\
\frac{\sum_{x=Y}^{M}{\Pr}_{9}(0,x)}{\sum_{x=Y}^{M}{\Pr}_{10}(x)},&M_{a}^{(t)}=0,M_{a}^{(t-1)}>0,\\
\frac{\sum_{x=Y}^{M}{\Pr}_{9}(x,0)}{{\Pr}_{10}(0)}, &M_{a}^{(t)}>0,M_{a}^{(t-1)}=0,\\
\frac{\sum_{x,y\in\{Y,\cdots,M\}}{\Pr}_{9}(x,y)}{\sum_{x=Y}^{M}{\Pr}_{10}(x)},&M_{a}^{(t)}>0,M_{a}^{(t-1)}>0.
\end{cases}
\label{eq:P6m}
\end{equation}

\subsection{Case S$_1$N$_0$B$_0$ (Active Channel Sensing, Channel Blocking, Fixed Channel Selection)}
\label{sec:cases1n0b0}

In this case the SU performs the coarse sensing for only one channel currently utilized for data transmission. Thus, the PU on a channel not used by the SU does not affect the OSA network, and as a result, the sensing stage of the OSA network becomes more sensitive to the location of the PUs in the radio spectrum. Therefore, probabilities related to the sensing stage $S$, such as ${\Pr}_{12}(\cdot)$ of (\ref{eq:P12}) and ${\Pr}_{15}(\cdot)$ of (\ref{eq:P15}) in Appendix~\ref{sec:p15s0n1b1}, and ${\Pr}_{20}(\cdot)$ of (\ref{eq:P20}) in Appendix~\ref{sec:P13S0N0B1} need to be updated.

First, we update the definition of ${\Pr}_{20}(\cdot)$. In option S$_1$N$_0$B$_0$, even if more than one channel is detected as idle, only one channel is utilized for data transmission. Thus, we consider only the case when one channel was detected as idle even though there can be more channels detected as idle. Then, obviously if a OSA network detects an idle channel, i.e. $S<2$, the number of channels detected as idle is one. If $S=2$, the probability that no idle channel is detected is also one. Thus ${\Pr}_{20}(\cdot)$ can be modified as
\begin{equation}
{\Pr}_{20}(\cdot)\triangleq
\begin{cases}
1,&X_a=0,S=2\textrm{~or~}X_a=1,S<2,\\
0,&\textrm{otherwise}.\\
\end{cases}
\end{equation}
Next, we present the modification of ${\Pr}_{12}(\cdot)$ in (\ref{eq:P12}). This modification is based on the fact that the sensing stage at time $t$ can be affected by the number and the position of WPUs and the NPUs at time $t-1$. Thus, denoting ${\Pr}_{25}\left(S^{(t)} | U_{w}^{(t)}, U_{n}^{(t)},S^{(t-1)},U_{w}^{(t-1)}, U_{n}^{(t-1)}\right)$ as the conditional probability of the sensing stage at time $t$ given the number of NPUs and WPUs for times $t$ and $t-1$, ${\Pr}_{12}(\cdot)$ is modified as follows
\begin{align}
{\Pr}_{12}(\cdot)\triangleq\;&{\Pr}_{15}\left(S^{(t-1)}| U_{w}^{(t-1)}, U_{n}^{(t-1)}\right){\Pr}_{16}\left( U_{w}^{(t)}, U_{n}^{(t)}| U_{w}^{(t-1)}, U_{n}^{(t-1)}\right)\nonumber\\&\times{\Pr}_{17}\left( U_{w}^{(t-1)}, U_{n}^{(t-1)}\right){\Pr}_{25}\left(S^{(t)} | U_{w}^{(t)}, U_{n}^{(t)},S^{(t-1)},U_{w}^{(t-1)}, U_{n}^{(t-1)}\right).
\label{eq:P12a}
\end{align}

For the same reason as in case of ${\Pr}_{12}(\cdot)$, ${\Pr}_{15}(\cdot)$ in (\ref{eq:P15}) needs to be modified as well. The modification process is presented in Appendix~\ref{sec:P15S1N0B0}.

\section{Numerical Results}
\label{sec:numerical_results}

In this section we provide performance results for all considered options. We note that all analytical results were verified via simulations using a method of batch means for 90\% confidence interval. Each simulation run, with a warm-up period of 10000 network events, was divided into 100 batches, where each batch contained 10000 network events.

Due to many parameters considered we limit our numerical investigation to three most representative case studies. That is, we consider the impact of varying number of NPUs, the impact of varying PU activity and the impact on two-stage spectrum sensing design on the system throughput. Results are presented in Section~\ref{sec:npu_impact}, Section~\ref{sec:pu_activity_impact} and Section~\ref{sec:sensing_impact}, respectively.

\subsection{Calculation of Average Subchannel Capacity $C$}
\label{sec:Cp_capacity}

Before we proceed with the presentation we need to comment on the calculation of $C$. Following the IEEE 802.22 model~\cite{ieee80222} we consider the PHY capacity $C_p$ and MAC layer overhead $\xi$ separately, such that $C=(1-\xi)C_p$, where $C_p=460.8$\,kbps\footnote{Assumptions: 16-QAM modulation with 4 bits per OFDM symbol and 1/2 channel coding per subcarrier; for guard band for NPUs, six subchannels of the IEEE 802.22 represent one subchannel in the system model. Uplink and downlink are time division duplexed, where 16 symbols and 8 symbols are assigned to downlink and uplink, respectively.}. 

For the MAC layer overhead $\xi$ we consider the frame structure of the IEEE 802.22 such that one downlink OFDM symbol of all subchannels is assigned to a preamble and two downlink OFDMA symbols of all subchannels are assigned to management messages. Also considering errors on the subchannels, we assume a bit error rate of $10^{-6}$. Hence, we calculate the MAC layer overhead reduction factor as $1-\xi =0.8125$. Therefore the total subchannel capacity is $C= 374.4$\,kbps.

\subsection{Impact of Varying Number of NPUs on OS-OFDMA Design}
\label{sec:npu_impact}

Throughout this section we assume that $X=4$ channels are available and split into $Y=10$ subchannels. Frame length is set to $t_f=20$\,ms, which represents the length of two frames of the IEEE 802.22~\cite{ieee80222}, i.e. it represents the time SU device searches for OFDM preambles on a given channel and highly conservative value of a inter-frame sensing interval~\cite[Table 233]{ieee80222}. False alarm and detection probability for the coarse sensing case is $\delta_a=0.99$ and $\phi_{a}=0.1$, respectively, while for the fine sensing case $\delta_a=1$ and $\phi_{a}=0$, respectively. Sensing time during coarse sensing is $\tau_{a}=0$. Since we have assumed that uplink and downlink are divided by time division duplex and coarse sensing is performed in the uplink, as in~\cite{ieee80222}, the sensing overhead is zero. In the fine sensing phase $\tau_f=\frac{3}{5}t_f$, modeling 3 consecutive frames used for the fine sensing and the following 2 frames used for data transmission until next sensing period. Furthermore we assume that $l_c=1$, $l_w=10$, $l_n=1$ subchannels are allocated to CBR, NPU and WPU connection, respectively. Note that the number of subchannels assigned to VBR connection, $l_v$, is varying depending on the network state. The maximum number of users of each class is $U_{v,\max}=2$, $U_{c,\max}=10$ and $U_{w,\max}=2$ (investigating impact of NPUs), and  $U_{v,\max}=2$, $U_{c,\max}=10$ and $U_{n,\max}=10$ (investigating impact of WPUs). Those values correspond to a small network and allow for an easier understanding of the subsequent numerical results.

Also, for the purpose of this section we set up inter-arrival and departure times taking into consideration the IEEE 802.22 network, where many active licensed users operate over the TV band. Since in general, the traffic pattern of PUs for such case is not well known (more discussion on this aspect is presented in Section~\ref{sec:pu_activity_impact}), for the WPU we keep the wireless assist video devices~\cite{reihl_report_2006} in mind, which can be assumed to broadcast on average four hours of signal transmission for every twelve hours on average in this scenario. For the NPU, we consider environment with numerous wireless microphones and assume that they appear every two hours on average and utilize channels for one hour on average. For CBR, considering voice or video transmission, we assume that on average five minute long CBR connection is generated for every five minutes on average. For VBR, we assume that data traffic is generated every two hours on average and continues for two hours if one channel is assigned for a VBR connection. For simulation efficiency, since we operate in large parameter ranges, we scale them down by setting the CBR connection arrival rate to one second with preserving the ratios between all traffic parameters, i.e. we normalize average inter-arrival and departure times of all users in the unit of five minutes by dividing them by 300 seconds. Thus for the large number of users, the inter-arrival time becomes shorter. For the analysis, we calculate the inter-arrival time by dividing the individual inter-arrival time by the maximum number of users. In summary, $1/\lambda_w=144/U_{w,\max}$ s, $1/\lambda_n=24/U_{n,\max}$\,s, $1/\lambda_c=1/U_{c,\max}$ s, $1/\lambda_v=12/U_{v,\max}$ s, $1/\mu_w=48$ s, $1/\mu_n=12$ s, $1/\mu_c=1$ s and $1/\mu_v=240$ s.

\begin{figure}
\centering
\subfigure[]{\includegraphics[width=0.32\columnwidth]{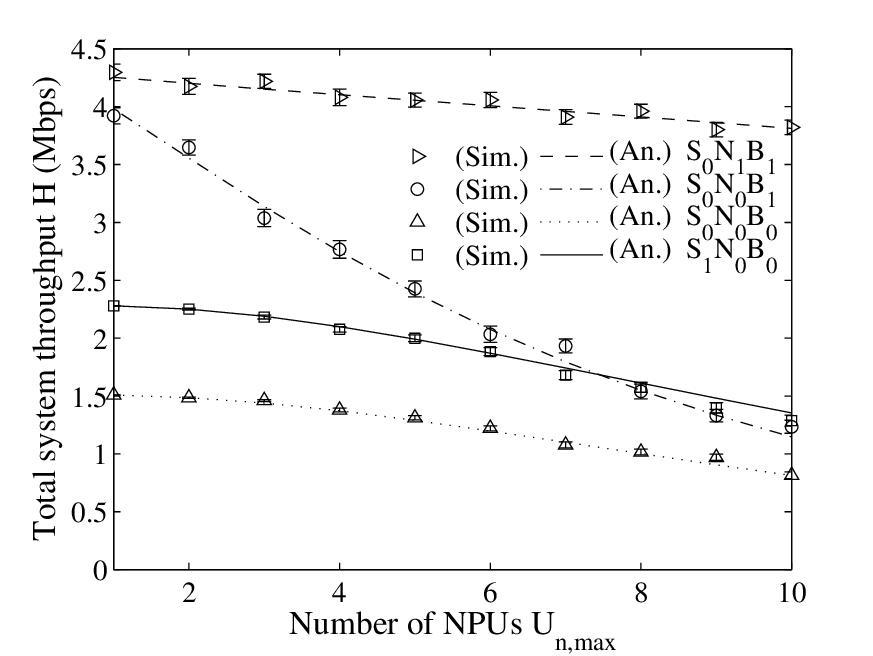}\label{fig:total_NPU}}
\subfigure[]{\includegraphics[width=0.32\columnwidth]{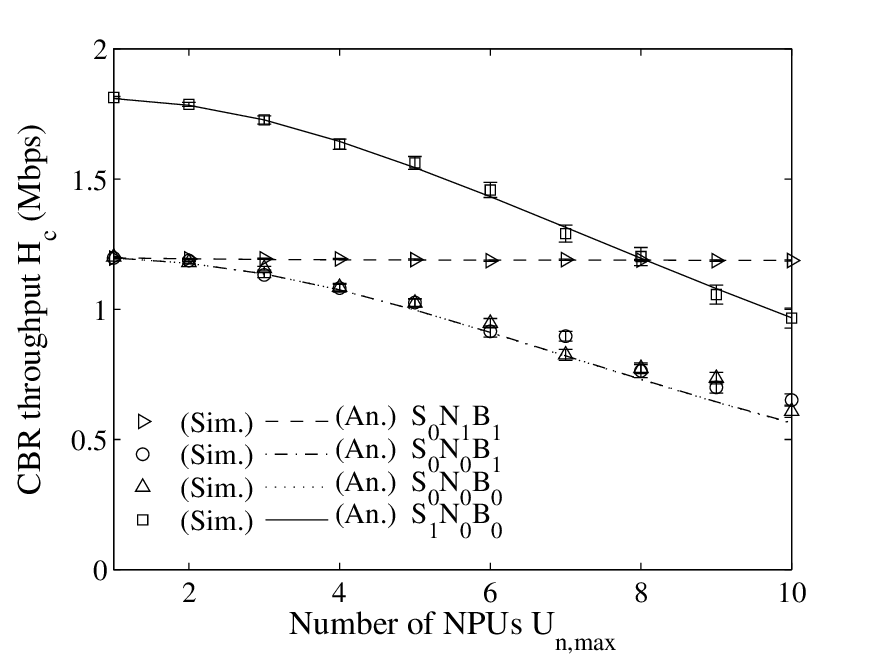}\label{fig:CBR_NPU}}
\subfigure[]{\includegraphics[width=0.32\columnwidth]{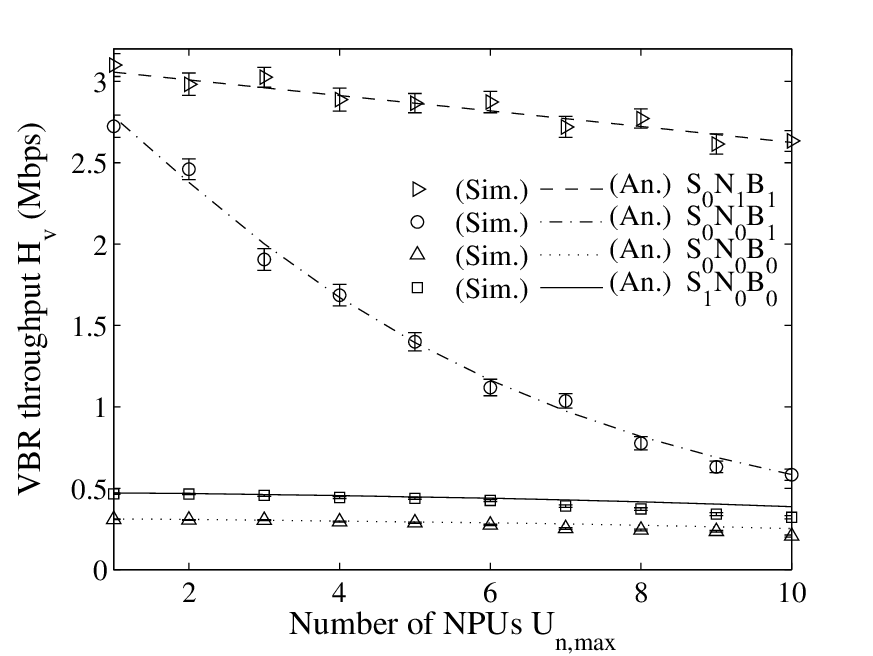}\label{fig:VBR_NPU}}\\
\subfigure[]{\includegraphics[width=0.32\columnwidth]{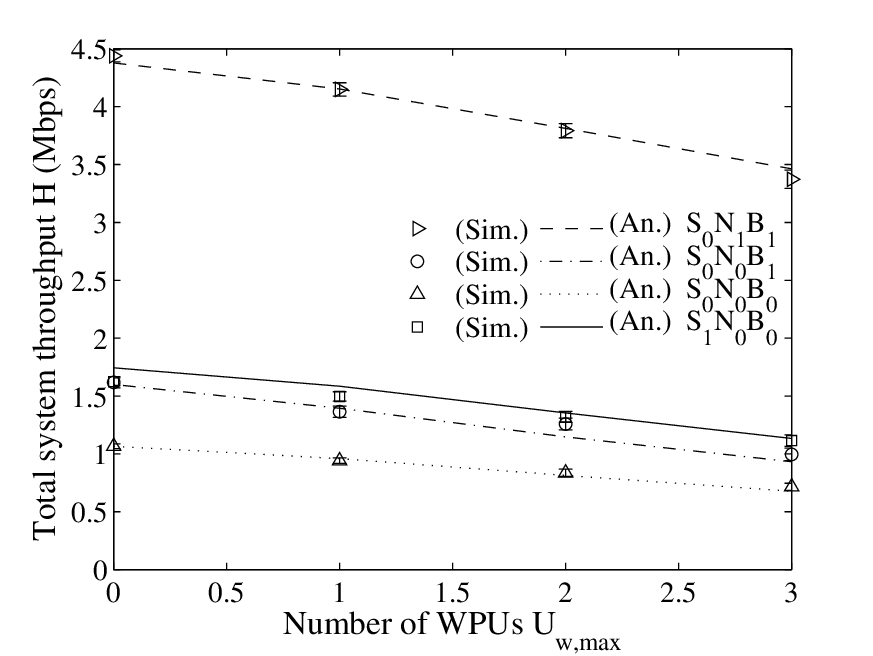}\label{fig:total_WPU}}
\subfigure[]{\includegraphics[width=0.32\columnwidth]{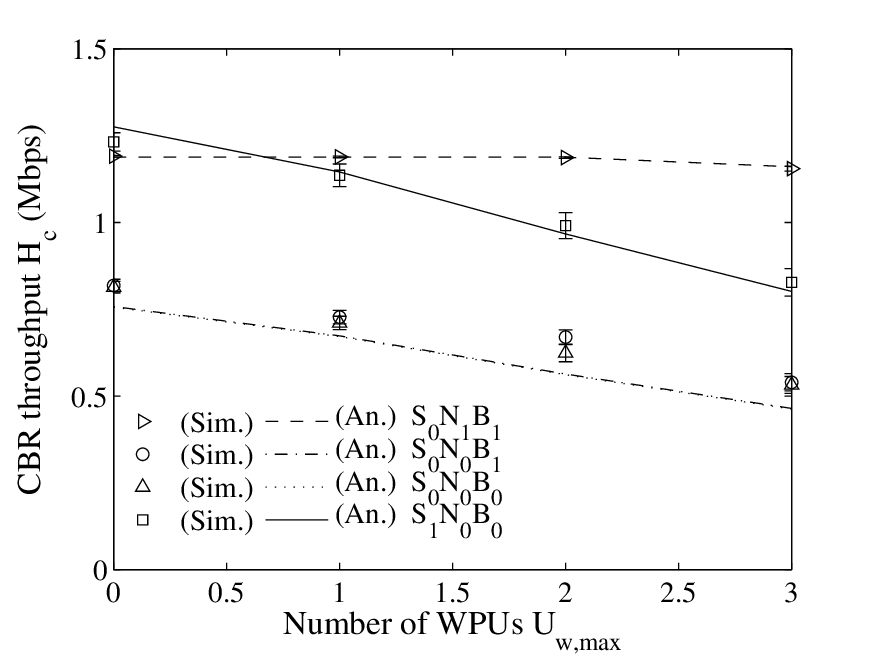}\label{fig:CBR_WPU}}
\subfigure[]{\includegraphics[width=0.32\columnwidth]{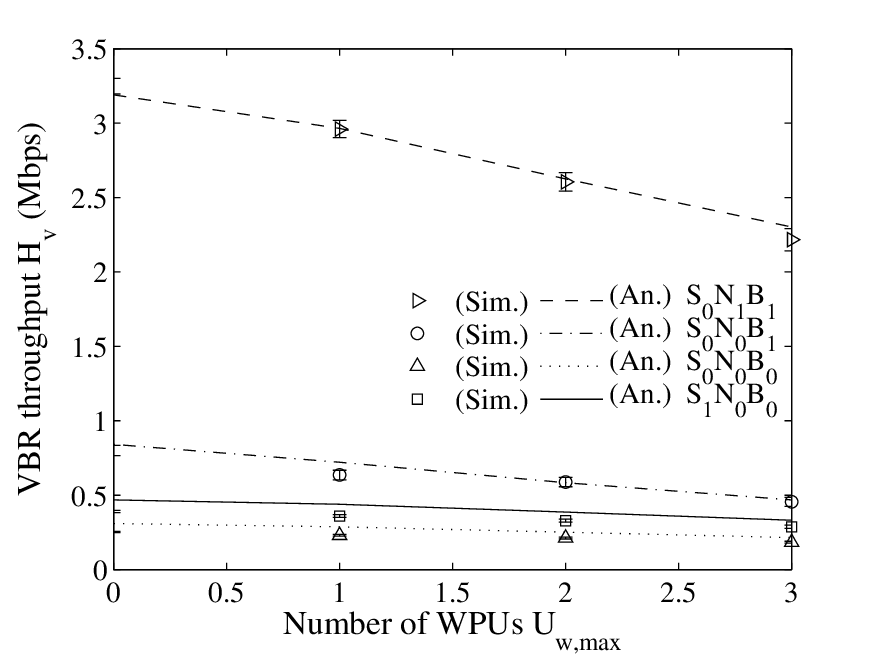}\label{fig:VBR_WPU}}
\caption{Performance of different OS-OFDMA options as a function of NPU, Fig.~(a), (b), (c), and WPU, Fig.~(d), (e), (f) for $X=4$, $Y=10$, $M=40$, $t_f=20$\,ms, $\delta_a=0.99$, $\phi_{a}=0.1$, $\tau_{a}=0$, $\delta_f=1$, $\phi_f=0$, $\tau_f=\frac{3}{5} t_f$, $C=374.4$\,kbps, $U_{w,\max}=2$ (WPU),  $U_{n,\max}=10$ (NPU), $l_w=10$, $l_n=1$, $U_{c,\max}=10$, $l_c=1$, $U_{v,\max}=2$; (a) (d) total system throughput, (b) (e) throughput of CBR connections, and (c) (f) throughput of VBR connections.}
\label{fig:npu}
\end{figure}

The results are presented in Fig.~\ref{fig:npu}. The throughput of every OS-OFDMA design option decreases with increasing numbers of NPUs and WPUs. First, we observe that S$_0$N$_1$B$_1$ is the best design option when total and VBR throughput is concerned, which is due to the highest flexibility in exploiting all spectrum opportunities. Second, interestingly with low number of NPUs and WPUs CBR throughput is higher for S$_1$N$_0$B$_0$ than for S$_0$N$_1$B$_1$, while with the high number of WPU and NPU the opposite holds, see Fig.~\ref{fig:CBR_NPU} and Fig.~\ref{fig:CBR_WPU}. This is because of the sensing overhead of the fine sensing that is performed more frequently for S$_0$N$_1$B$_1$ than for S$_1$N$_0$B$_0$. Third, there is no difference in CBR throughput between S$_0$N$_0$B$_1$ and S$_0$N$_0$B$_0$ when the number of NPUs varies. This is because in this network setup CBR, which has higher priority than the VBR, utilizes enough resources even though channel bonding is not applied. Fourth, the S$_0$N$_0$B$_1$ design option is extremely sensitive to the activity of the NPUs. The throughput of this design option decreases rapidly as the number of NPUs increases, compare Fig.~\ref{fig:total_NPU} with Fig.~\ref{fig:VBR_NPU}. This effect is not visible however when the NPU number is kept fixed, but the number of WPUs changes, see Fig.~\ref{fig:total_WPU} and Fig.~\ref{fig:VBR_WPU} and compare with Fig.~\ref{fig:total_NPU} and Fig.~\ref{fig:VBR_NPU}, respectively. This is because S$_0$N$_0$B$_1$ is sensitive to the position of NPUs in the spectrum, i.e. the lack of subchannel notching (N$_0$), causes this option to perform worse. 

Further, we demonstrate a benefit of active channel sensing strategy in two-stage sensing. Surprisingly, the total throughput of S$_1$N$_0$B$_0$ is greater than that of S$_0$N$_0$B$_1$ for large number of NPUs and WPUs, see Fig.~\ref{fig:total_NPU} and Fig.~\ref{fig:total_WPU}, even though the implementation for S$_0$N$_0$B$_1$ enables wider bandwidth sensing than S$_1$N$_0$B$_0$. This is because for S$_0$N$_0$B$_1$, due to channel bonding, the probability that OSA network needs to sense the channel is much higher and network needs to perform sensing often. While for S$_1$N$_0$B$_0$ only actively used channel is sensed.

Note that the simulation results agree with the analysis in all figures. A slight mismatch between simulation and analysis for cases S$_0$N$_0$B$_1$, S$_0$N$_0$B$_0$ and S$_1$N$_0$B$_0$ in Fig.~\ref{fig:CBR_WPU} is a result of the approximation used in calculating throughput for these design options, see (\ref{eq:P24}), (\ref{eq:P25S}). More discussion on this issue is presented in Section~\ref{sec:pu_activity_impact}.

\subsection{Impact of PU Activity on OS-OFDMA Design}
\label{sec:pu_activity_impact}

To investigate the impact of varying PU activity on the performance of different OS-OFDMA designs, for ease of explanation, we have considered to focus on NPUs only. Then, as a case study, we consider wireless microphones as an example of NPU. The parameters of the OS-OFDMA are the same as in Section~\ref{sec:npu_impact}. Before presenting the performance results we need to estimate the most realistic values of wireless microphones activity descriptors, i.e. average arrival rate and channel occupancy time.

In the case of average NPU channel occupancy time we set it to a value between one and four hours, believing this represents a common activity time. More discussion is needed, however, on the arrival rate of the wireless microphones. Since the potential number of wireless microphones is location dependent, we have setup four different network scenarios, representing different places in the USA, see Table~\ref{tab:network_scenarios}, that differ in population density $\rho$ and activity time $1/\mu_n$. We assume that OS-OFDMA BS covers a fraction of the area of diameter $L=2$\,mi (for all scenarios) of the considered location, while the wireless microphones move in and out of the BS circular coverage with a certain speed $v=1.5$\,mph (for all scenarios). Then using a fluid flow model approximation~\cite{lam_commag_1997} we compute the average crossing rate of the wireless microphones to that area and translate it directly to an average arrival rate of wireless microphone on any of the subchannels. That is $\lambda_n=\rho h\pi Lv$, where $1/h$ denotes number of inhabitants per one active wireless microphone in the considered location. Since the value of $h$ is not known reliably\footnote{The only credible report we were able to find was~\cite{etsitr102546}. The estimation using data present in this report was based on a simple calculation. According to~\cite[Sec. A2]{etsitr102546} there were 1924431 wireless microphone shipments in the European Union (EU) between 2002 and 2006, which translates to $\approx$ 1 wireless microphone per 1000 EU inhabitants (assuming a constant level of wireless microphone shipments per year). Note that the value of 35,000--70,000 licensed wireless microphone operations in USA presented in~\cite{chen_jsac_2011} was not substantiated with any reference.} we assume that one wireless microphone is present per 300 inhabitants (for all scenarios) and such wireless microphone is active for 10\% of the time. Finally, $U_{n,\max}=\max(\lfloor\pi(L/2)^2\rho h\rfloor,1)$ in this case. We set the inter-arrival and departure times for other users as the same value as Section~\ref{sec:npu_impact}. Also for analysis, we normalize all time parameters in the unit of five minutes as in Section~\ref{sec:npu_impact}.
\begin{table}
\centering
\caption{Scenarios for the Analysis of NPU Activity Impact on the Performance of OS-OFDMA Designs}
\begin{tabular}{c||cccc}
\hline
Nickname & ``heavy urban'' & ``urban'' & ``light urban'' & ``event'' \\ 
Location in US & Los Angeles, CA & Santa Barbara, CA & Madison, WI & Staples Center, CA\\
Users/mi$^2$, $\rho$ & 7452.7 & 4708.2 & 2701.0 & 7452.7\\
Activity time, $1/\mu_n$\,h & 1 & 1 & 1 & 4\\
$U_{n,\max}$ & 8 & 5 & 3 & 18 \\
\hline
\end{tabular}
\label{tab:network_scenarios}
\end{table}

\begin{figure}
\centering
\subfigure[``heavy urban'']{\includegraphics[width=0.49\columnwidth]{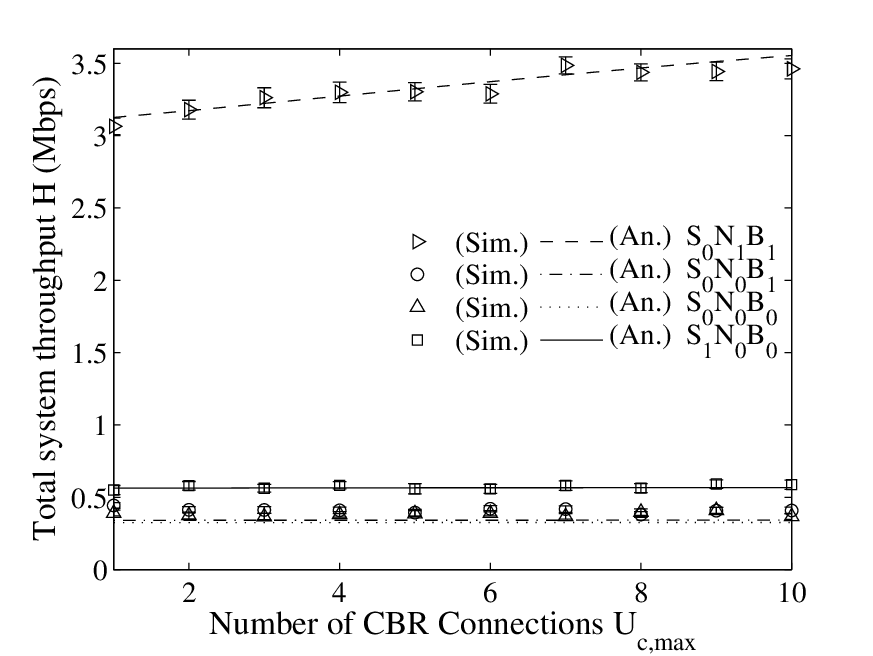}\label{fig:total_HeavyUrban}}
\subfigure[``urban'']{\includegraphics[width=0.49\columnwidth]{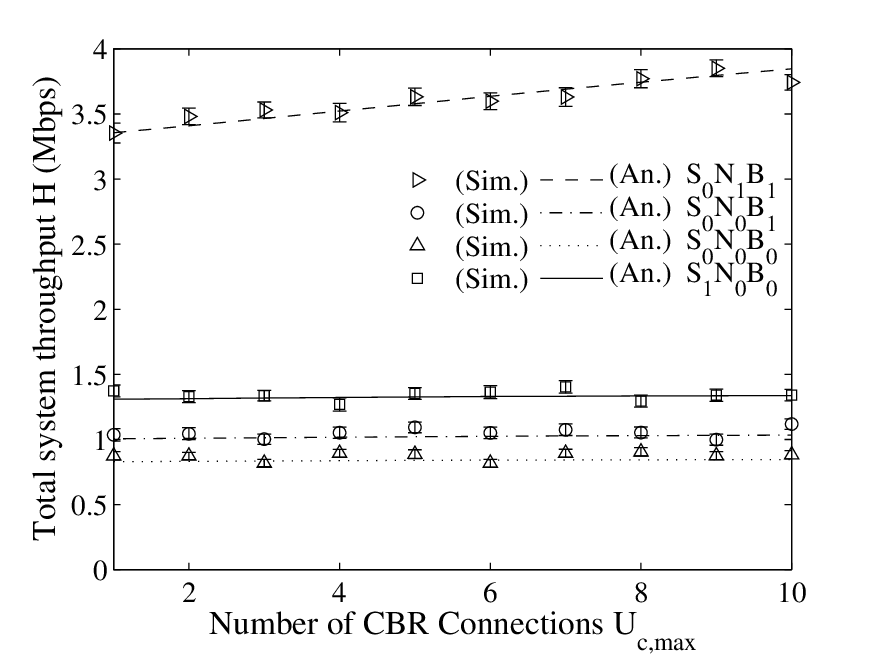}\label{fig:total_Suburban}}
\subfigure[``light urban'']{\includegraphics[width=0.49\columnwidth]{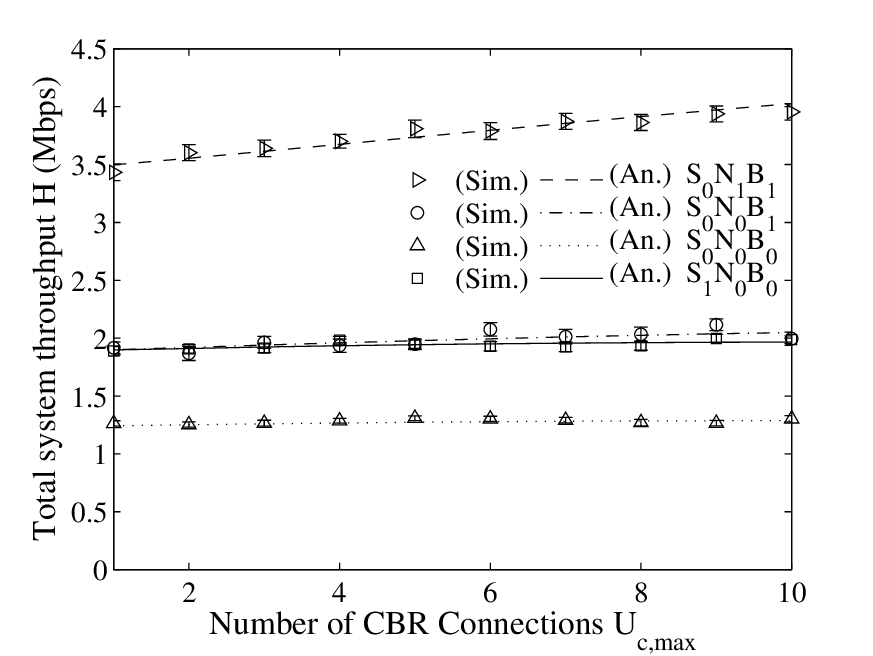}\label{fig:total_Rural}}
\subfigure[``event'']{\includegraphics[width=0.49\columnwidth]{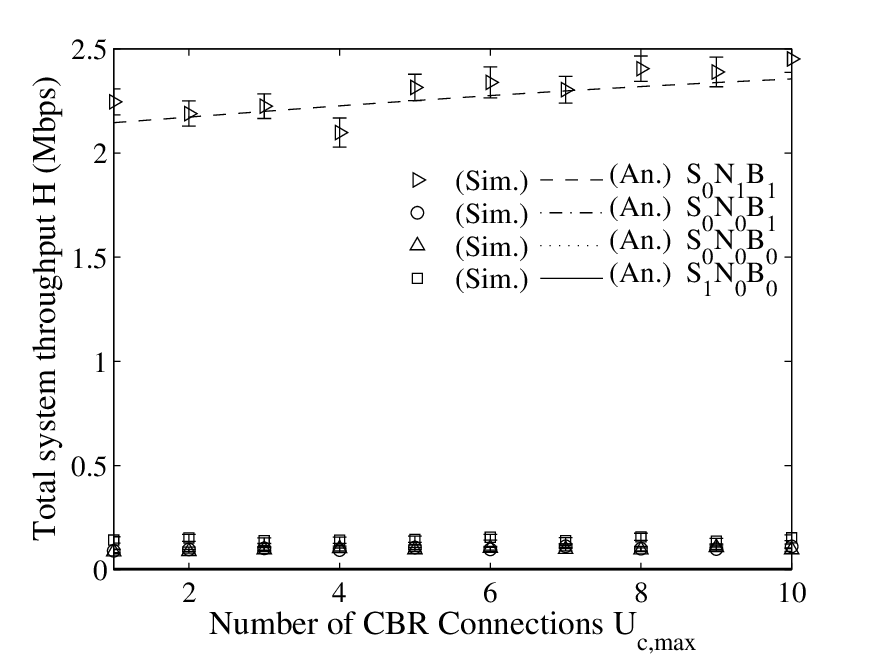}\label{fig:total_SE}}
\caption{Total throughput of different OS-OFDMA implementations for all network cases; Parameter setup and the ordering of the figures is the same as in Fig.~\ref{fig:npu}: (a) ``heavy urban'', (b) ``urban'', (c) ``light urban'', (d) ``event''.}
\label{fig:total_npu}
\end{figure}

\begin{figure}
\centering
\subfigure[``heavy urban'']{\includegraphics[width=0.49\columnwidth]{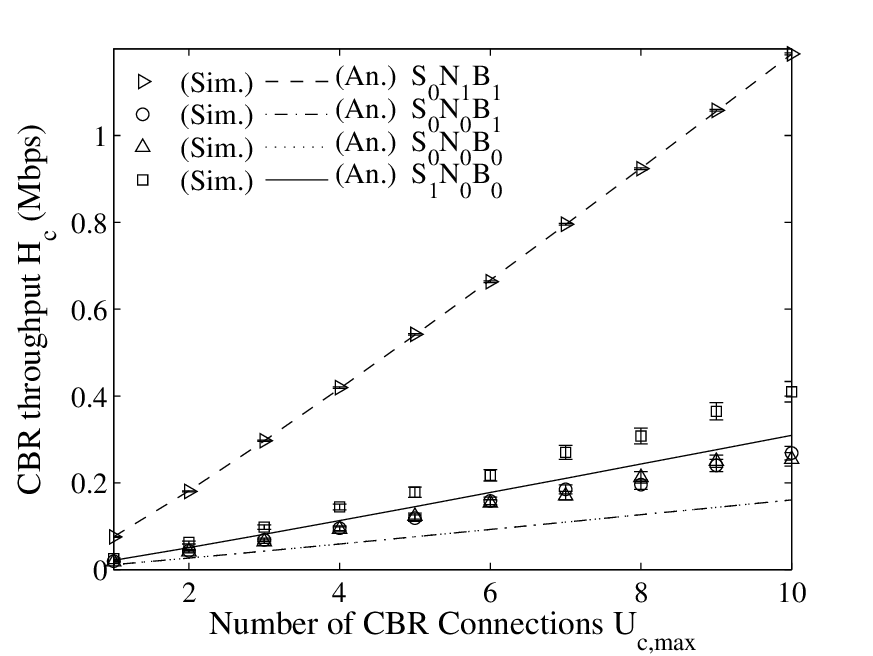}\label{fig:CBR_HeavyUrban}}
\subfigure[``urban'']{\includegraphics[width=0.49\columnwidth]{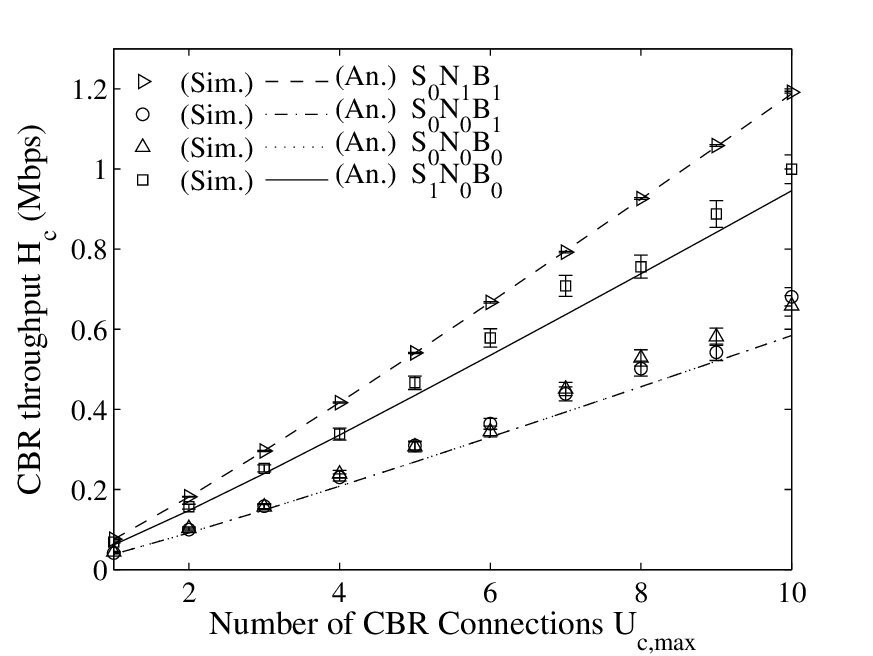}\label{fig:CBR_Suburban}}
\subfigure[``light urban'']{\includegraphics[width=0.49\columnwidth]{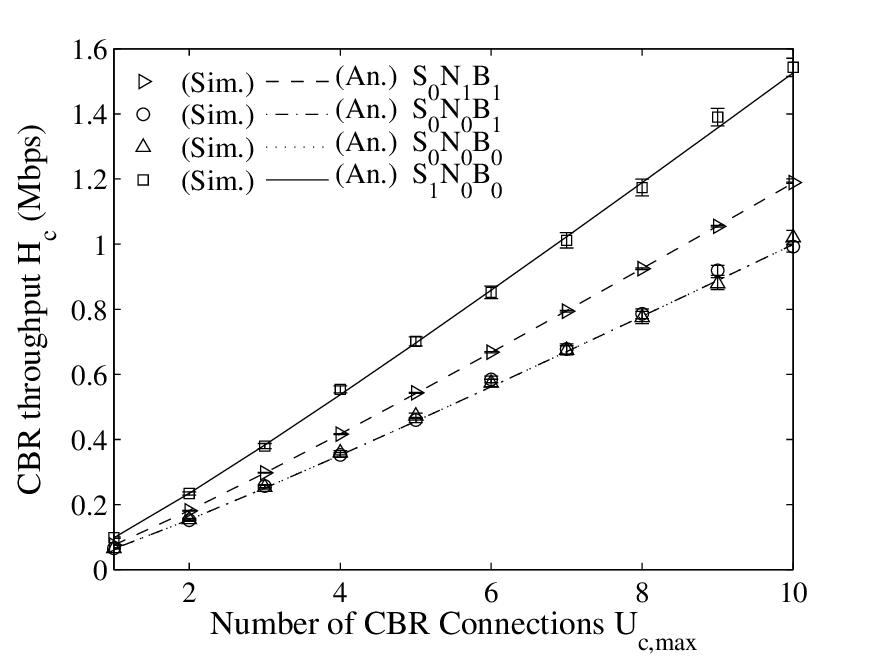}\label{fig:CBR_Rural}}
\subfigure[``event'']{\includegraphics[width=0.49\columnwidth]{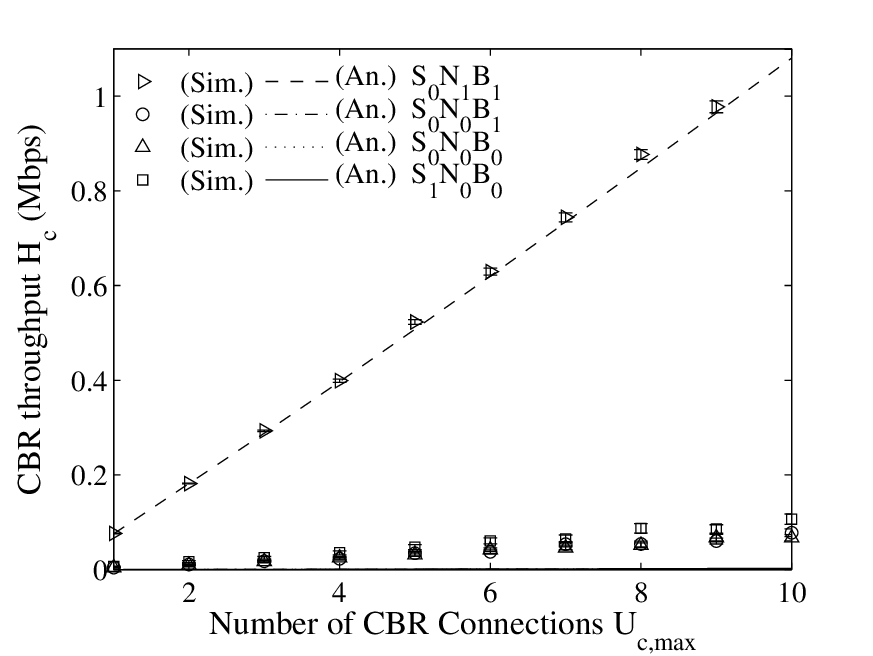}\label{fig:CBR_SE}}
\caption{CBR throughput of different OS-OFDMA implementations for all network cases. Parameter setup and the ordering of the figures is the same as in Fig.~\ref{fig:npu}.}
\label{fig:cbr_npu}
\end{figure}

\begin{figure}
\centering
\subfigure[``heavy urban'']{\includegraphics[width=0.49\columnwidth]{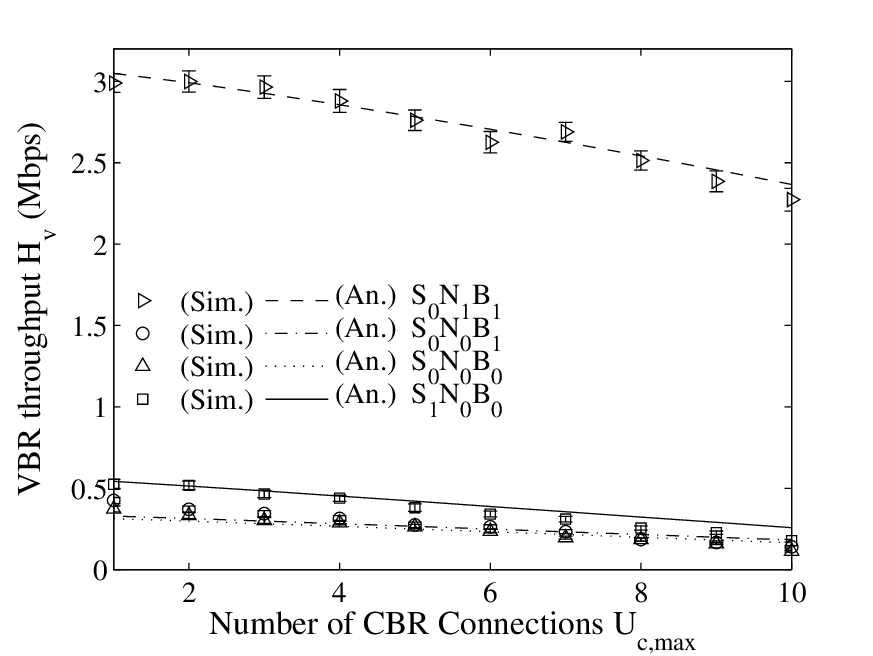}\label{fig:VBR_HeavyUrban}}
\subfigure[``urban'']{\includegraphics[width=0.49\columnwidth]{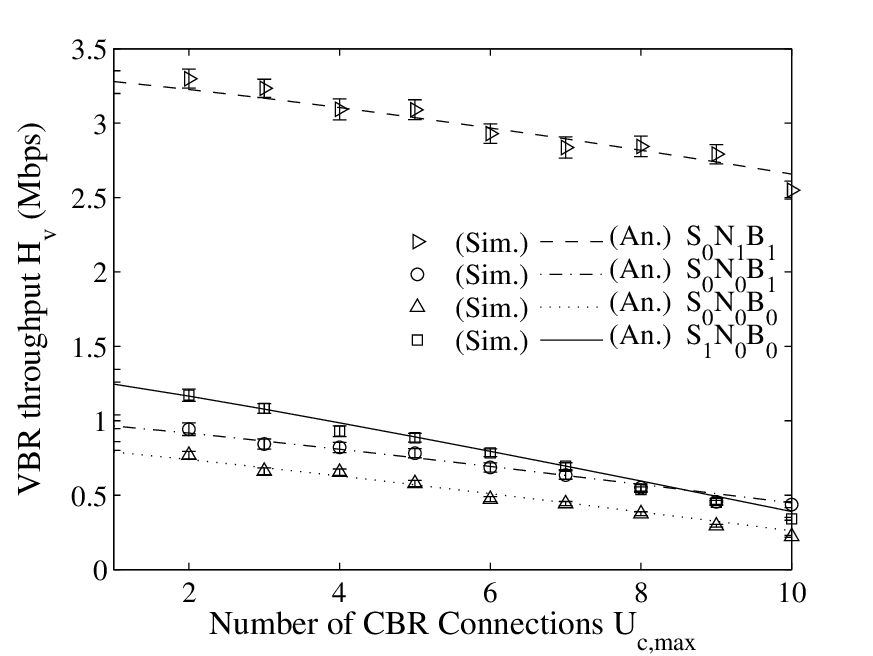}\label{fig:VBR_Suburban}}
\subfigure[``light urban'']{\includegraphics[width=0.49\columnwidth]{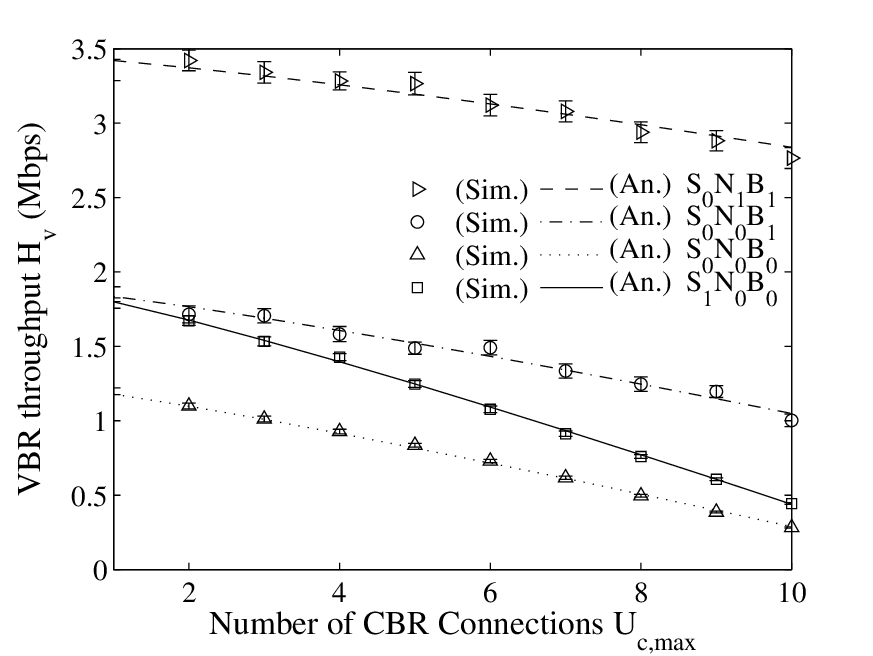}\label{fig:VBR_Rural}}
\subfigure[``event'']{\includegraphics[width=0.49\columnwidth]{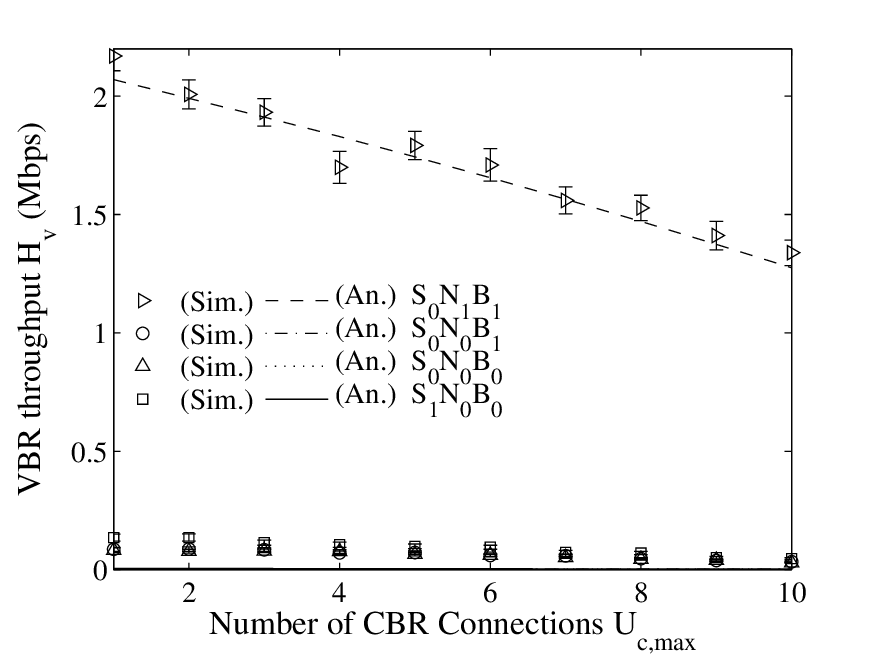}\label{fig:VBR_SE}}
\caption{VBR throughput of different OS-OFDMA implementations for all network cases. Parameter setup and the ordering of the figures is the same as in Fig.~\ref{fig:npu}.}
\label{fig:vbr_npu}
\end{figure}

The results are grouped separately for total average network throughput, CBR throughput and VBR throughput, see Fig.~\ref{fig:total_npu}, Fig.~\ref{fig:cbr_npu} and Fig.~\ref{fig:vbr_npu}, respectively. We have chosen to vary number of CBR connections in all figures as a parameter, since in our model CBR connection is the most QoS sensitive and capacity demanding SU traffic class. First we immediately observe that S$_0$N$_1$B$_1$ implementation obtains the highest throughput. The larger the activity of the wireless microphones, the bigger the difference between S$_0$N$_1$B$_1$ and the remaining implementations -- compare for example Fig.~\ref{fig:total_Rural} and Fig.~\ref{fig:total_SE}. Option S$_0$N$_1$B$_1$ obtains a throughput around 3\,Mbps, even in the ``event'' scenario. This is due to maximum utilization of the remaining channel capacity by subchannel notching and channel bonding. As the NPU activity increases all implementations reach almost zero throughput, while S$_0$N$_1$B$_1$ still obtains reasonable performance. The worst performance is obtained for S$_0$N$_0$B$_0$, while the S$_1$N$_0$B$_0$ and S$_0$N$_0$B$_1$ are in between the extreme cases. As the activity of the wireless microphones decreases all implementations start to converge in throughput -- compare for example Fig.~\ref{fig:CBR_HeavyUrban} with Fig.~\ref{fig:CBR_Rural}, however the individual relation between the implementations stays the same. Then, we observe that all implementations except S$_0$N$_1$B$_1$ obtain a similar throughput, irrespective of the network scenario. This proves that subcarrier notching promises to deliver most of the available capacity in the licensed bands. 

The worst situation, in terms of network scenario, is the ``event'' scenario. Due to long channel dwell time by NPU, i.e. four hours, the throughput for all implementations except S$_0$N$_1$B$_1$ reaches zero. We also conclude that in scenarios where the activity of the wireless microphones is low, like in the ``light urban'' scenario, the users of systems based on OS-OFDMA are promised to obtain high QoS, see Fig.~\ref{fig:total_Rural}, Fig.~\ref{fig:CBR_Rural} and Fig.~\ref{fig:VBR_Rural}, irrespective of the implementation.

A separate comment is needed for simulation verification of the results. In all cases S$_0$N$_1$B$_1$ implementation matches simulations perfectly, irrespective of the parameters selected. However, due to the approximations assumed for the remaining OS-OFDMA implementations, see again (\ref{eq:P24}), (\ref{eq:P25S}), the slight discrepancy is particularly visible for the scenarios with high NPU activity rate, compare for example Fig.~\ref{fig:CBR_HeavyUrban} with Fig.~\ref{fig:CBR_Rural}. The results prove, on the other hand, that the developed model works very well for low PU activities, which is the typical case in real life PU occupancy statistics~\cite{wellens_phycom_2009,staple_spectrum_2004}. Still, for each case study the relation between each OS-OFDMA implementation is well captured for any value of the parameters considered, while for the majority of the cases the mismatch between simulations and analysis is less than 10\%.

\subsection{Impact of Two-stage Spectrum Sensing Options on OS-OFDMA Design}
\label{sec:sensing_impact}

The final experiment considers effect of sensing design parameters on the performance of OS-OFDMA designs. For this investigation we change the sensing time of coarse sensing, which has a direct effect on false alarm probability and makes a significant impact on the frequency of switching to fine sensing (and as a result on the throughput of the system). We keep other parameters the same as in the first experiment described in Section~\ref{sec:npu_impact}, except $U_{n,\max}=2$, to see the effect of coarse sensing clearer, and $U_{c,\max}=10$. We do not alter parameters of fine sensing phase, since we want to explore the benefit of two-stage sensing and the coarse sensing phase is a common element of every sensing method, including single stage sensing.

Considering Rayleigh Channel with Additive White Gaussian Noise, we compute false alarm probability, $p_{10}$, and detection probability, $p_{11}$, for an individual SU user as~\cite[Eq. (3)]{Park_arxiv_2009} and~\cite[Eq. (4)]{Park_arxiv_2009}, respectively. Then, according to~\cite[Sec. III-B]{Park_arxiv_2009}, we can derive $p_{11}$ as a function of $p_{10}$ and $\tau_a$ for given average PU SNR and subchannel bandwidth $b$. Assuming collaborative sensing of all CBR and VBR users and OR logic for combining scheme, we compute system false alarm probability as
\begin{equation}
\phi_a=1-(1-p_{10})^{(U_{c,\max}+U_{v,\max})},
\label{eq:psi_a}
\end{equation}
and system detection probability, $\delta_a$, as (\ref{eq:psi_a}) replacing $p_{10}$ with $p_{11}$. In this evaluation we keep the detection probability $\delta_a=0.99$ and change the sensing time for the coarse sensing such that $\tau_a=(0,4)$\,ms. Note that $\tau_a=0$\,ms represents single stage sensing. Table~\ref{tab:sens_setup} presents calculated false alarm probability, $\phi_a$, based on the assumed sensing time $\tau_a$.

\begin{table}
\centering
\caption{Sensing Time and the Respective False Alarm Probability for the Experiment Introduced in Section~\ref{sec:sensing_impact}}
\begin{tabular}{c||c|c|c|c|c|c|c|c|c}
\hline
$\tau_a$ (ms) & 0 & 0.5 & 1 & 1.5 & 2 & 2.5 & 3 & 3.5 & 4\\
\hline
$\phi_a$ & 1 & 0.2308 & 0.0446 & 0.0087 & 0.0018 & 0.0004 & 0.0001 & 1.57e-4 & 0.333e-4\\
\hline
\end{tabular}
\label{tab:sens_setup}
\end{table}

\begin{figure}
\centering
\subfigure[]{\includegraphics[width=0.32\columnwidth]{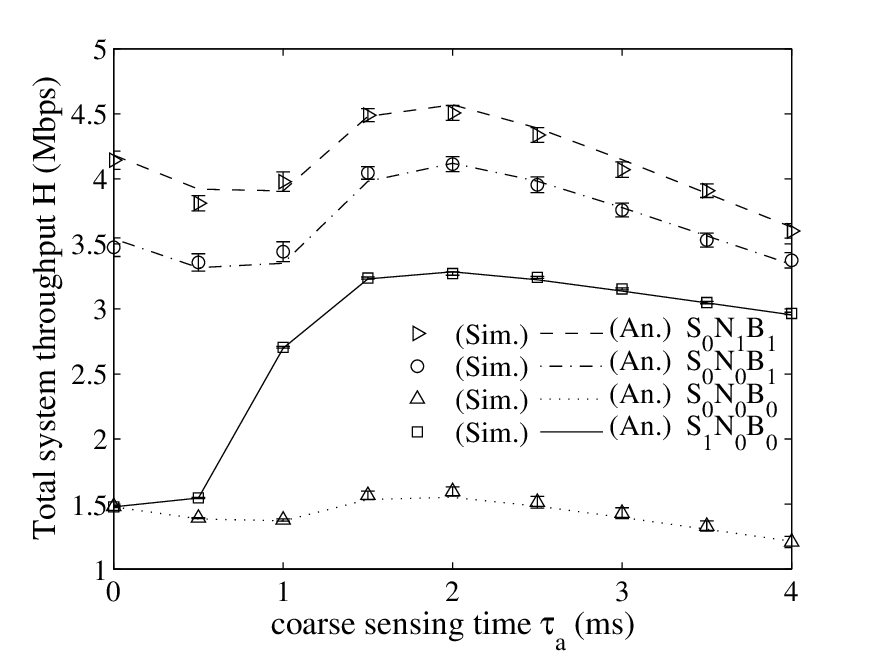}\label{fig:total_ts}}
\subfigure[]{\includegraphics[width=0.32\columnwidth]{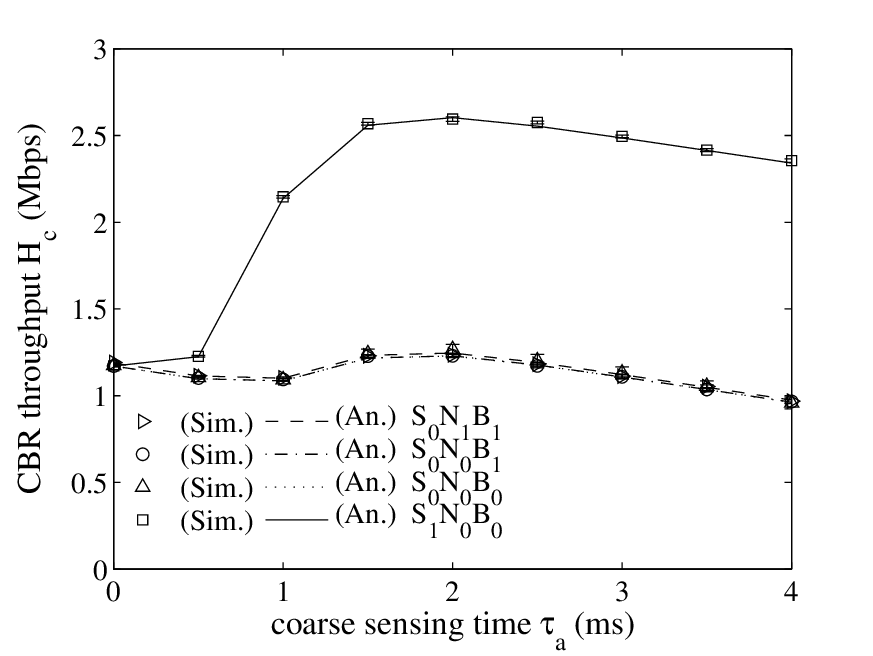}\label{fig:CBR_ts}}
\subfigure[]{\includegraphics[width=0.32\columnwidth]{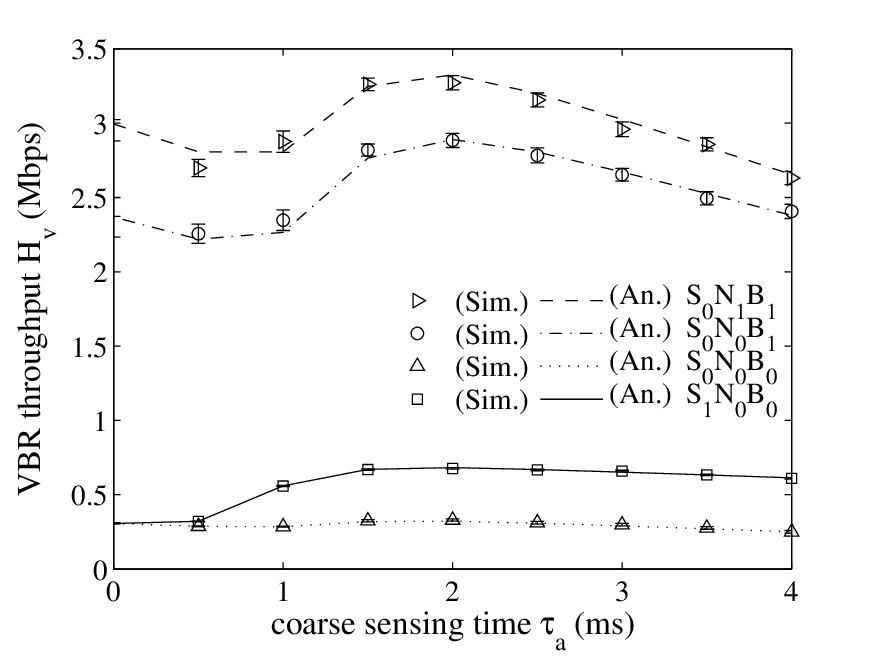}\label{fig:VBR_ts}}
\caption{Throughput of OS-OFDMA implementations, (a) total, (b) CBR, and (c) VBR, as a function of coarse sensing time. Parameter setup is the same as in Fig.~\ref{fig:npu}, except for $U_{n\max}=2$, $U_{c,\max}=10$ while $\tau_a$ and $\phi_a$ are given in Table~\ref{tab:sens_setup}.}
\label{fig:sensing}
\end{figure}

The results are presented in Fig.~\ref{fig:sensing}. First interesting observation is that two stage sensing does not always provide better performance than single stage sensing. Comparing the total throughput at $\tau_a=0$ with $\tau_a\approxeq(0.5,1)$\,ms of all OS-OFDMA options, two stage sensing shows worse performance than single stage sensing. This result is due to high probability of false alarm for this range of $\tau_a$, see Table~\ref{tab:sens_setup}. In addition, for this network setup, the throughput is maximized at $\tau_a=2$\,ms and is larger than for $\tau_a=0$ which confirms that two-stage sensing can indeed benefit all OS-OFDMA operations. This also confirms that the design choice of the IEEE 802.22 for spectrum sensing method was correct. When $\tau_a$ increases beyond the point for which the throughput is largest, the throughput of all OS-OFDMA implementations starts to rapidly decrease. This is due to the fact that the sensing overhead starts to dominate over potential improvement from decreased false alarm rate. This so called sensing-throughput tradeoff is in agreement with a similar investigation in the context of OSA ad hoc networks~\cite{Park_arxiv_2009}. Note that the relation between sensing time and the obtained throughput are the same when looking at total, CBR and VBR throughput, see the shapes of all curves in Fig.~\ref{fig:total_ts}, Fig.~\ref{fig:CBR_ts} and Fig.~\ref{fig:VBR_ts}. Also, the different order in obtained throughput for each OS-OFDMA implementation are due to the specific OS-OFDMA options, not due to spectrum sensing parameters selected. For that, compare, e.g., the position of S$_1$N$_0$B$_0$ in Fig.~\ref{fig:total_ts} with the position of the same implementation in Fig.~\ref{fig:CBR_ts}. For more on this aspect please refer to Section~\ref{sec:pu_activity_impact}. Finally, note that the simulation results match well with the analysis for all OS-OFDMA implementations.

\section{Conclusions}
\label{sec:conclusions}

In this paper we have proposed an analytical model that allows the comparison of different designs of OS-OFDMA. We have considered design options that include channel bonding, subchannel notching and two stage spectrum sensing. As a performance metric we have derived average throughput obtained by the secondary users of the spectrum. In the analysis we have included the inter-relations of different connection classes and priorities, like constant and variable bit rate traffic of the secondary users, and wideband and narrowband primary users. We concluded that OS-OFDMA design that allows the flexible bonding of channels and notching of subchannels currently occupied by the primary users, obtains the highest throughput in comparison to the designs that do not consider those options. As one of our numerical results show, the improvement reaches couple of hundred percent when the activity level of the different primary user types is very high. Also, as our investigation show, the two-stage spectrum sensing technique used in OS-OFDMA proves to increase the average network throughput, provided the probability of false alarm in the coarse sensing stage is low.

\appendices

\section{Derivation of $G_x(k|U_x,\lambda_x)$ and $T_x(j|U_x,\mu_x)$}
\label{app:TG}

If the inter-arrival time has a negative exponential distribution with average arrival rate $\lambda_x$, the number of connections generated in a frame of length $t_f$ has a Poisson distribution. However, because we limit the maximum number of users to $U_{x,\max}$, the number of users including newly generated connections, $U_x$, cannot exceed $U_{x,\max}$. Considering this, we derive the probability of $k$ new connections being generated in a frame, $G_x(k|U_x,\lambda_x)$, as
\begin{equation}
G_x(k|U_x,\lambda_x)\triangleq\\
\begin{cases}
\frac{(\lambda_x t_f)^{k}e^{-\lambda_x t_f}}{k!},&k \geq 0, U_x<U_{x,\max};\\
\sum\limits_{i=k}^{\infty} \frac{(\lambda_x t_f)^{i}e^{-\lambda_x t_f}}{i!},&k \geq 0, U_x=U_{x,\max};\\
0,&k<0.
\end{cases}
\label{eq:G}
\end{equation}
Note that the subscript $x=\{w,n,c,v\}$ indicates the class of users, i.e., $w$ for WPU, $n$ for NPU, $c$ for CBR, and $v$ for VBR.

Further, denoting the departure rate of each connection as $\mu_x$, the probability that $j$ connections are released from $U_x$ active connections during a frame of length $t_f$, $T_x(j|U_x,\mu_x,t_f)$, can be calculated recursively as
\begin{equation}
T_x(j|U_x,\mu_x,t_f)\triangleq
\begin{cases}
\int_0^{t_f}U_x \mu_x e^{-U_x \mu_x t} T_x(j-1|U_x-1,\mu_x,t_f-t) dt,& j>0;\\
e^{-U_x \mu_x t_f},& j=0,
\end{cases}
\label{eq:T}
\end{equation}
which after some manipulation reduces to
\begin{equation}
T_x(j|U_x,\mu_x,t_f)={U_x \choose j} e^{-U_x \mu_x t_f}(e^{\mu_x t_f}-1)^j.
\end{equation}
Since we only consider the connection release probability for users within the duration of a frame we abbreviate $T_x(j|U_x,\mu_x)\triangleq T_x\left(j|U_{x},\mu_x,t_f\right)$.

\section{Derivation of $\Pr_{13}(\cdot)$ for S$_0$N$_1$B$_1$}
\label{sec:p13s0n1b1}

In $\Pr_{13}(\cdot)$ in (\ref{eq:P11}), $M_a$ subchannels include $M_m$ subchannels that are occupied by PUs but mis-detected and $M_0$ subchannels correctly detected as idle. Therefore, $M_a=M_m+M_0$, and thus to compute $\Pr_{13}(\cdot)$ we define a supporting probability, $\Pr_{14}(M_{m},M_{0}|U_{w},U_{n},S)$, which is the probability that the number of subchannels detected as idle correctly and falsely are $M_0$ and $M_m$, respectively. 

In design option S$_0$, a SU performs the coarse sensing for all subchannels first, and then, if the SU detects a PU on a subchannel, it immediately switches to fine sensing and senses all subchannels again with high sensing accuracy. Thus, for a condition $S=0$ (only coarse sensing is performed), the number of subchannels detected as idle must be the same as the number of all subchannels, i.e. $M_m+M_0=M$. In other words, the case $M_m+M_0<M$ is impossible for the condition $S=0$. For the condition $S=2$ (no idle subchannel is detected after all stages of sensing) only the case $M_m+M_0=0$ is possible. For $S=1$, using the detection probability $\delta_f$ and the false alarm probability $\phi_f$ in the fine sensing stage, we can derive the probability that $M_m$ busy subchannels are mis-detected and $M_0$ idle subchannels are correctly detected for given $U_w$ WPUs and $U_n$ NPUs. Thus,
\begin{equation}
{\Pr}_{14}(\cdot)\triangleq
\begin{cases}
F & \begin{split}S=1\end{split};\\
1, & \begin{split}S=0, M_m+M_0=M, \text{~or~} S=2, M_m+M_0=0; \end{split}\\
0, & \begin{split}S=0, M_m+M_0<M, \text{~or~} S=2, M_m+M_0>0, \end{split}
\end{cases}
\label{eq:P14}
\end{equation}
where $F=\binom{M_{p}}{M_m}(1-\delta_{f})^{M_m}\delta_{f}^{M_{p}-M_m}\binom{M-M_{p}}{M_0} (1-\phi_{f})^{M_0}\phi_{f}^{M-M_{p}-M_0},$ and $M_{p}=\min(M,U_w l_w + U_n l_n)$ is the number of subchannels actually occupied by PUs. Therefore, $\Pr_{13}(\cdot)$ can be derived as
\begin{equation}
{\Pr}_{13}(\cdot)\triangleq\sum_{x=0}^{M_a}{\Pr}_{14}(x,M_a-x|U_{w},U_{n},S).
\label{eq:P13}
\end{equation}

\section{Derivation of $\Pr_{15}(\cdot)$ for S$_0$N$_1$B$_1$}
\label{sec:p15s0n1b1}

To derive $\Pr_{15}(\cdot)$ we need to consider the following three cases. First, if the OSA network mis-detects existing PUs and correctly detects all idle subchannels in the coarse sensing stage, $S=0$ because OSA network will not advance to fine sensing. Second, if the OSA network detects at least one PU correctly or falsely in the coarse sensing stage and detects all subchannels as busy also correctly or falsely in the fine sensing stage, $S=2$ because after coarse and fine sensing no idle subchannel is detected. Otherwise $S=1$. Thus,
\begin{equation}
{\Pr}_{15}(\cdot)\triangleq
\begin{cases}
(1-\delta_a)^{M_p} (1-\phi_{a})^{M-M_p}, & S=0;\\
\left(1-(1-\delta_a)^{M_p} (1-\phi_{a})^{M-M_p}\right)\delta_f^{M_p} \phi_f^{M-M_p}, & S=2;\\
1-{\Pr}_{15}(S=0|U_{w},U_{n})-{\Pr}_{15}(S=2|U_{w},U_{n}), & S=1.
\end{cases}
\label{eq:P15}
\end{equation}

\section{Derivation of $\Pr_{13}(\cdot)$ for S$_0$N$_0$B$_1$}
\label{sec:P13S0N0B1}

For the case S$_0$N$_0$B$_1$, because the spectrum is utilized based on the unit of bandwidth of one channel rather than one subchannel, we introduce the probability of the number of channels detected as idle, $X_a$, to compute the probability of the number of subchannels detected as idle, $M_a$. Denoting ${\Pr}_{20}(X_a|U_w,U_n,S)$ as the probability that the number of channels detected as idle is $X_a$ given $U_w$ WPUs and $U_n$ NPUs in the $S$ sensing stage, (\ref{eq:P13}) is modified as
\begin{equation}
{\Pr}_{13}(\cdot)\triangleq 
\begin{cases}
{\Pr}_{20}(X_a|U_w,U_n,S),&M_{a}=X_a Y;\\
0,&\text{otherwise}.
\end{cases}
\label{eq:P13a}
\end{equation}
There are two required conditions for ${\Pr}_{20}(\cdot)$ in (\ref{eq:P13a}). First, the sum of the channels detected as idle and the number of WPUs cannot be greater than the total number of channels, i.e. $X_a+U_w \leq X$. Second, the number of NPUs cannot be greater than the total number of subchannels that are not occupied by the WPUs, i.e. $U_n \leq (X-U_w)\lfloor Y/l_n \rfloor$. We denote $\mathbf{1}_{c}$ as an indicator of those conditions, defining $\mathbf{1}_{c}=1$ for $X_a+U_t \leq X$ or $U_n \leq (X-U_w)\lfloor Y/l_n \rfloor$ and $\mathbf{1}_{c}=0$, otherwise.

Next, we compute ${\Pr}_{20}(\cdot)$ under the feasible conditions considering different sensing stages. Under the condition that no channel is detected as idle after the fine sensing, i.e. $S=2$, the number of channels detected as idle is $X_a=0$ under the assumption of perfect PU detection. On the other hand, when all channels are detected as idle in the coarse sensing, i.e. $S=0$, only $X_a=X$ is possible. The analysis for the condition that the SU performs fine sensing and detects idle channels, i.e. $S=1$, is not easy to derive directly because the number of channels detected as idle depends on the position of NPUs in the spectrum as well as the number of NPUs. Thus, defining the number of channels actually occupied by the NPU as $X_n$, we deconstruct ${\Pr}_{20}(\cdot)$ for $S=1$ into two components: the conditional probability of $X_n$ for given number of PUs and the sensing stage, denoted as ${\Pr}_{21}(X_n|U_w,U_n,S)$, and the conditional probability of the number of channels detected as idle for a given $X_n$, denoted as ${\Pr}_{22}(X_a|X_n,U_w,U_n,S)$. Then,
\begin{equation}
{\Pr}_{20}(\cdot)\triangleq 
\begin{cases}
\sum_{\mathcal{X}} {\Pr}_{21}(X_n|U_w,U_n,S){\Pr}_{22}(X_a|X_n,U_w,U_n,S),&\begin{split}\mathbf{1}_{c}=1, S=2;\end{split}\\
1,&\begin{split}&\mathbf{1}_{c}=1\text{\,and\,}S=0,\\ &X_a=X\text{\,or\,}S=2,\\ &X_a=0;\end{split}\\
0,&\begin{split}\textrm{otherwise},\end{split}\\
\end{cases}
\label{eq:P20}
\end{equation}
where $\mathcal{X}=\{X_n|\lceil (U_n l_n /Y\rceil\leq X_n \leq {\min}(U_n,X-U_t-X_a)\}$ because $X_n$ is the smallest when all NPUs are located on adjacent subchannels, i.e. $\lceil (U_n l_n)/Y\rceil$, and the largest when all NPUs are located in different channels separated as far as possible, i.e. ${\min}(U_n,X-U_t-X_a)$.

To compute ${\Pr}_{21}(\cdot)$ in (\ref{eq:P20}), we assume that any NPU can appear on any subchannel with equal probability, and that false alarm can occur uniformly over all idle subchannels. Then, we introduce a supporting function $f_s(k,x,r)$ denoting the number of possibilities that $k$ items are distributed over exactly $x$ bins each of which has a capacity of $r$ items. 

To derive $f_s(k,x,r)$ first we introduce the supporting variable, $i_j$ -- the number of items in $j$-th bin where $j \in \{1,\cdots,x\}$. Then, there can be $\binom{r}{i_j}$ possible distributions for the $j$-th bin, and thus for $x$ bins there can be $\sum_{j=1}^{i_1} \binom{r}{j} \cdots \sum_{j=1}^{i_x} \binom{r}{j}$ possibilities. Because there should be no bin empty, $i_j \geq 1$. Also, each of the bins has a capacity of $r$ items and the total number of items cannot be greater than $k$, and therefore $i_j \leq \min(r,k)$. In addition, if the number of bins is less than the number of items or equal to zero, there is no way to fill all $x$ bins. Thus
\begin{equation}
f_s(k,x,r)=
\begin{cases}
\sum_{\mathcal{I}(k,x,r)}\sum_{j=1}^{i_1} \binom{r}{j} \cdots \sum_{j=1}^{i_x} \binom{r}{j},&0<x<k;\\
0,\!\!\!\!&\text{otherwise},
\end{cases}
\label{eq:fs}
\end{equation}
where $\mathcal{I}(k,x,r)=\{i_1,\cdots,i_x|\sum_{j=1}^{x}{i_j} =k,i_1,\cdots,i_x \in \{1,\cdots,\min(r,k)\} \}$.

Then ${\Pr}_{21}(\cdot)$ can be computed by dividing the number of possible events that $U_n$ NPUs are located on exactly $X_n$ channels each of which can have at maximum $\lfloor Y/l_n \rfloor$ NPUs, i.e. $f_s(U_n,X_n,\lfloor Y/l_n \rfloor)$, by the number of all possible events that $U_n$ NPUs appear on $X-U_w$ channels, i.e. $\binom{(X-U_w)\lfloor Y/l_n \rfloor}{U_n}$. Note that if there is no NPU, i.e. $U_n=0$, then $X_n$ should be zero. Considering the possible case of selecting $X_n$ NPU channels from a total of $X-U_w$ channels, i.e. $\binom{X-U_w}{X_n}$, we derive ${\Pr}_{21}(\cdot)$ as
\begin{equation}
{\Pr}_{21}(\cdot)=
\begin{cases}
1,& X_n=0, U_n=0,\\
\frac{\binom{X-U_w}{X_n} f_s(U_n,X_n,\lfloor Y/l_n \rfloor)} {\binom{(X-U_w)\lfloor Y/l_n \rfloor}{U_n}},& \text{otherwise}.
\end{cases}
\label{eq:P21}
\end{equation}

Now we present the derivation of ${\Pr}_{22}(\cdot)$ in (\ref{eq:P20}). If there exist $U_w$ and $X_n$ channels that are occupied by WPUs and NPUs, respectively, and $X_a$ channels are correctly detected as idle, the remaining $X-U_w-X_n-X_a$ channels must be falsely detected as busy. Considering the number of events of selecting $X_a$ channels from $X-U_w-X_n$ idle channels, we derive ${\Pr}_{22}(\cdot)$ as
\begin{align}
{\Pr}_{22}(\cdot)=&\:\binom{X-U_w-X_n}{X_a} (1-(1-\phi_f)^Y)^{X-U_w-X_n-X_a}(1-\phi_f)^{YX_a}.
\end{align}

\section{Derivation of $\Pr_{23}(\cdot)$ for for S$_0$N$_0$B$_1$}
\label{sec:P23S0N0B1}

Because for S$_0$N$_0$B$_1$ the spectrum is utilized based on the unit of bandwidth of one channel, denoting a supporting probability ${\Pr}_{24}\left(X_{a}^{(t)}| X_{a}^{(t-1)}, U_{w}^{(t)}, U_{w}^{(t-1)}, U_{n}^{(t)}, U_{n}^{(t-1)}, S^{(t)}, S^{(t-1)}\right)$ as the conditional probability of the number of channels, we derive ${\Pr}_{23}(\cdot)$ as follows
\begin{align}
{\Pr}_{23}(\cdot)\triangleq
\begin{cases} 
{\Pr}_{24}(X_{a}^{(t)}|X_{a}^{(t-1)},U_{w}^{(t)}, U_{w}^{(t-1)}, U_{n}^{(t)}, U_{n}^{(t-1)}, S^{(t)}, S^{(t-1)}),&M_{a}=X_a Y;\\ 
0,&\text{otherwise}. 
\end{cases}
\label{eq:P23}
\end{align}
To reduce the complexity of calculating ${\Pr}_{24}(\cdot)$ we use an approximation that if the number of PUs is the same at times $t$ and $t-1$, the positions of NPUs at times $t$ and $t-1$ are also the same. This approximation is valid when the PU activity is not high. With this approximation we ignore the case that a certain number of NPUs disappear while at the same time the same number of new NPUs appear, but at different locations. Also we assume that if the number of WPUs and NPUs have changed, the number of channels detected as idle is independent of the number of WPUs and NPUs at time $t-1$, which means that ${\Pr}_{24}(\cdot)\approxeq{\Pr}_{20}(\cdot)$. Considering these approximations we derive ${\Pr}_{24}(\cdot)$ as
\begin{equation}
{\Pr}_{24}(\cdot)\approxeq
\begin{cases}
1,&\begin{split}&U_{w}^{(t)}=U_{w}^{(t-1)},U_{n}^{(t)}=U_{n}^{(t-1)},X_a^{(t)}=X_a^{(t-1)};\end{split}\\
0,&\begin{split}&U_{w}^{(t)}=U_{w}^{(t-1)},U_{n}^{(t)}=U_{n}^{(t-1)},X_a^{(t)}\neq X_a^{(t-1)};\end{split}\\
{\Pr}_{20}(X_a|U_w,U_n,S),&\begin{split}\textrm{otherwise},\end{split}
\end{cases}
\label{eq:P24}
\end{equation}
where $\Pr_{20}(\cdot)$ is defined in Appendix~\ref{sec:P13S0N0B1} as (\ref{eq:P20}).

\section{Derivation of ${\Pr}_{15}(\cdot)$ for S$_0$N$_0$B$_1$}
\label{sec:P15S0N0B1}

To modify ${\Pr}_{15}(\cdot)$ in (\ref{eq:P15}), under the approximation of the perfect detection, depending on the number of subchannels occupied by PUs, i.e. $M_p$, we consider three cases. In the first case, $M_p=M$, all subchannels are occupied by PUs. Then, because of the perfect detection approximation $S=2$ is the only feasible condition. In the second case, $M_p=0$, no PU appears on the subchannels. In this case, the false alarm probability of each sensing stage affects the probability ${\Pr}_{15}(\cdot)$. In the last case, $0<M_p<M$, the probability of $S=0$ is zero and the false alarm probability of the coarse sensing does not affect the performance. Thus
\begin{equation}
{\Pr}_{15}(\cdot)=
\begin{cases}
1,&M_p=M, S=2;\\
(1-\phi_{a})^M,&M_p=0, S=0;\\
1-(1-\phi_{a})^M-\phi_{a}^M (\phi_f)^M,&M_p=0,S=1;\\
\phi_{a}^M \phi_f^M,&M_p=0,S=2;\\
1-\phi_f^{M-M_p},&0<M_p<M,S=1;\\
\phi_f^{M-M_p},&0<M_p<M,S=2;\\
0,&\textrm{otherwise}.
\end{cases}
\label{eq:P15m}
\end{equation}

\section{Derivation of $\Pr_{15}(\cdot)$ for S$_1$N$_0$B$_0$}
\label{sec:P15S1N0B0}

Following the same convention as for the derivation of (\ref{eq:P6}) in Section~\ref{sec:s0n1b1}, denoting ${\Pr}_{26}(S,U_w,U_n)$ as the joint probability of the number of PUs and the sensing stage, we derive ${\Pr}_{15}(\cdot)$ as
\begin{equation}
{\Pr}_{15}(\cdot)\triangleq\frac{{\Pr}_{26}(S,U_w,U_n)}{{\Pr}_{17}(U_w,U_n)}.
\label{eq:P15a}
\end{equation}
Because expression ${\Pr}_{26}(\cdot)$ is the steady state probability, denoting the state transition probability as ${\Pr}_{27}\left(S^{(t)},U_w^{(t)},U_n^{(t)}|S^{(t-1)},U_w^{(t-1)},U_n^{(t-1)}\right)$, ${\Pr}_{26}(\cdot)$ can be obtained by solving the Markov chain with a state $\{S,U_w,U_n\}$, such that ${\Pr}_{26}(\cdot)=\sum_{\mathcal{S}^{(t-1)}}{\Pr}_{26}(\cdot){\Pr}_{27}(\cdot)$ and $\sum_{\mathcal{S}}{\Pr}_{26}(\cdot)=1$, where $\mathcal{S}$ is the set of all possible $S$, $U_{w}$, and $U_{n}$, and $\mathcal{S}^{(t-1)}$ is the set of the same variables at time $t-1$. Then ${\Pr}_{27}(\cdot)$ is computed as
\begin{align}
{\Pr}_{27}(\cdot)\triangleq{\Pr}_{25}\left(S^{t} | U_{w}^{(t)}, U_{n}^{(t)},S^{(t-1)},U_{w}^{(t-1)}, U_{n}^{(t-1)}\right){\Pr}_{16}\left(U_{w}^{(t)}, U_{n}^{(t)}| U_{w}^{(t-1)}, U_{n}^{(t-1)}\right).
\label{eq:P27}
\end{align}

To compute ${\Pr}_{12}(\cdot)$ in (\ref{eq:P12a}) and ${\Pr}_{15}(\cdot)$ in (\ref{eq:P15a}), we need to derive ${\Pr}_{25}(\cdot)$. For the derivation of ${\Pr}_{25}(\cdot)$, to avoid prohibitive complexity of calculation, we consider two specific cases. The first case is that there is no change for the number of WPUs and NPUs between time $t-1$ and $t$. This case is expected to happen most frequently because the PU generally has longer inter-arrival time and packet length than the frame size of the SUs. The second case is that the numbers of WPUs and NPUs are reduced as time goes from $t-1$ to $t$. If the PU activity is not too high, this case happens when some of the existing PUs disappear and no new PU enters into the subchannels. Then subchannels detected as idle at time $t-1$ still remains idle even at time $t$, which can affect the sensing stage significantly. For the other cases, we use a general approach. Depending on the changes of the numbers of the PUs, ${\Pr}_{25}(\cdot)$ is derived as
\begin{equation}
{\Pr}_{25}(\cdot)\approxeq
\begin{cases}
{\Pr}_{25a}\left(D\right),&U_w^{(t)}=U_w^{(t-1)},U_n^{(t)}=U_n^{(t-1)},\\
{\Pr}_{25b}\left(D\right),&U_w^{(t)}\leq U_w^{(t-1)},U_n^{(t)} \leq U_n^{(t-1)},\\
{\Pr}_{25c}\left(D\right),&\textrm{otherwise},
\end{cases}
\label{eq:P25S}
\end{equation}
where $D=\{S^{(t)} | U_{w}^{(t)}, U_{n}^{(t)},S^{(t-1)},U_{w}^{(t-1)}, U_{n}^{(t-1)}\}$. To analyze the first case in (\ref{eq:P25S}), where there is no change in the number of WPUs and NPUs between time $t-1$ and $t$, we reuse the assumption that there is no change in the positions at subchannels of the NPUs for the same number of NPUs at times $t$ and $t-1$, see Section~\ref{sec:s0n0b1}. With this approximation, under the condition that there exists at least one channel at time $t-1$, i.e. $S^{(t-1)}<2$, there should also exist at least one idle channel at time $t$, i.e. $S^{(t)}<2$. Thus, if no false alarm occurs, only coarse sensing is performed, i.e. $S^{(t)}=0$ and otherwise $S^{(t)}=1$. Next, for the condition that no channel is detected as idle at time $t-1$, i.e. $S^{(t-1)}=2$, there can be a large number of possible events depending on the positions of NPUs and false alarms. Thus, instead of considering all the possible events, we consider two cases and apply approximations for each one. The first case is that there exists a small number of NPUs at time $t-1$ so that the NPUs can not occupy all $X-U_w^{(t-1)}$ channels, i.e. $f_s\left(U_n^{(t-1)},X-U_w^{(t-1)},\lfloor Y/l_n \rfloor\right)=0$, where $f_s(\cdot)$ is defined in Appendix~\ref{sec:P13S0N0B1} as (\ref{eq:fs}). In this case, because there exists at least one idle channel, we approximate that the SU performs fine sensing and detects the idle channel at time $t$, i.e. $S^{(t)}=1$. In contrast, for the second case where there exists enough NPUs such that they are spread over all $X-U_w^{(t-1)}$ channels, i.e. $f_s\left(U_n^{(t-1)},X-U_w^{(t-1)},\lfloor Y/l_n \rfloor\right)>0$, we approximate that all channels are detected as busy at time $t$, i.e. $S^{(t)}=2$. Considering all these sub-cases, we derive ${\Pr}_{25a}(\cdot)$ as
\begin{equation}
{\Pr}_{25a}(\cdot)\triangleq
\begin{cases}
(1-\phi_{a})^Y,&\begin{split}S^{(t)}=0,S^{(t-1)}<2;\end{split}\\
1-(1-\phi_{a})^Y,&\begin{split}S^{(t)}=1,S^{(t-1)}<2;\end{split}\\
1,& \begin{split}S^{(t)}=2,J^{(t-1)}> 0,\textrm{~or~}S^{(t)}=1,J^{(t-1)}=0; \end{split}\\
0,&\textrm{otherwise,}
\end{cases}
\label{eq:P25b}
\end{equation}
where $J^{(x)}=f_s\left(U_n^{(x)},X-U_w^{(x)},\lfloor Y/l_n \rfloor\right)$. Next, we analyze the second case in (\ref{eq:P25S}), i.e. $U_w^{(t)}\leq U_w^{(t-1)}$ and $U_n^{(t)}\leq U_n^{(t-1)}$, which implies that the number of PUs decreases. Similar to the first case in (\ref{eq:P25S}), if there exist idle channels at time $t-1$, i.e. $S^{(t-1)}<2$, there must also exist idle channels at time $t$, i.e. $S^{(t)}<2$. If a false alarm does not occur on any of the $Y$ subchannels in a channel utilized for data transmission, the SU performs only coarse sensing, i.e. $S^{(t)}=0$, and otherwise, $S^{(t)}=1$. On the other hand, for the condition that there is no idle channel at time $t-1$, i.e. $S^{(t-1)}=2$, the probability of the sensing stage at time $t$ is calculated by dividing the number of the cases that $U_n^{(t)}$ NPUs occupy exactly all $X-U_w^{(t)}$ channels, $f_s\left(U_n^{(t)},X-U_w^{(t)},\lfloor Y/l_n \rfloor\right)$, by the number of all possible cases $\binom{(X-U_w^{(t)}) \lfloor Y/l_n \rfloor}{U_n^{(t)}}$. Note that when $S^{(t-1)}=2$, it is not possible to perform only the coarse sensing at time $t$, because the channel that is going to be utilized for data transmission is not determined. In other words, the probability of $S^{(t-1)}=0$ under $S^{(t-1)}=2$ equals zero. As a result, the probability of $S^{(t-1)}=1$ under $S^{(t-1)}=2$ equals to one minus the probability of $S^{(t-1)}=2$ under $S^{(t-1)}=2$. Thus, we derive ${\Pr}_{25b}(\cdot)$ as
\begin{equation}
{\Pr}_{25b}(\cdot)\triangleq
\begin{cases}
(1-\phi_{a})^Y,&S^{(t)}=0,S^{(t-1)}<2;\\
1-(1-\phi_{a})^Y,&S^{(t)}=1,S^{(t-1)}<2;\\
\frac{J^{(t)}}{\binom{(X-U_w^{(t)}) \lfloor Y/l_n \rfloor}{U_n^{(t)}}},&S^{(t)}=2,S^{(t-1)}=2;\\
1-\frac{J^{(t)}}{\binom{(X-U_w^{(t)}) \lfloor Y/l_n \rfloor}{U_n^{(t)}}},&S^{(t)}=1,S^{(t-1)}=2,\\
0,&\textrm{otherwise}.
\end{cases}
\label{eq:P25b}
\end{equation}

Finally, we compute ${\Pr}_{25c}(\cdot)$ in (\ref{eq:P25S}). In this case, again to simplify the computation, we ignore the effect of the sensing stage at time $t-1$ and only focus on the sensing stage at time $t$. Then, we compute the probability that the SU performs only coarse sensing, i.e. $S^{(t)}=0$, by dividing the number of cases that at least one channel remains idle and no false alarm occurs on that channel, $(1-\phi_{a})^Y\binom{(X-U_w^{(t)}-1) \lfloor Y/l_n \rfloor}{U_n^{(t)}}$, by all possible cases of all channels remaining idle, $\binom{(X-U_w^{(t)}) \lfloor Y/l_n \rfloor}{U_n^{(t)}}$. On the other hand, the probability that the SU detects no idle channel, i.e. $S^{(t)}=2$, is calculated by considering the case that $U_n$ NPUs occupy exactly all $X-U_w^{(t)}$ channels, i.e. $f_s(U_n^{t},X-U_w^t,\lfloor Y/l_n \rfloor)$. Note that for the case that there appears too many NPUs on the spectrum so that there is no possibility that even one channel cannot remain idle, only $S=2$ is feasible. This case can be denoted as $V= (X-U_w^{(t)}-1) \lfloor Y/l_n \rfloor$, because $X-U_w^{(t)}$ channels are available for the NPU and each channel can have a maximum $\lfloor Y/l_n \rfloor$ NPUs. Considering all these cases, we have
\begin{equation}
{\Pr}_{25c}(\cdot)\triangleq
\begin{cases}
\frac{(1-\phi_{a})^Y \binom{(X-U_w^{(t)}-1) \lfloor Y/l_n \rfloor}{U_n^{(t)}}}{\binom{(X-U_w^{(t)}) \lfloor Y/l_n \rfloor}{U_n^{(t)}}},&S^{(t)}=0, U_n^{(t)} \leq V;\\
\frac{J^{(t)}}{\binom{(X-U_w^{(t)}) \lfloor Y/l_n \rfloor}{U_n^{(t)}}}, &S^{(t)}=2, U_n^{(t)} \leq V;\\
1-{\Pr}_{25c}\left(S^{(t)}=0|\cdot\right)-{\Pr}_{25c}\left(S^{(t)}=2|\cdot\right),&S^{(t)}=1, U_n^{(t)} \leq V;\\
1,&S^{(t)}=2, U_n^{(t)} >V;\\
0,&\textrm{otherwise}.
\end{cases}
\label{eq:P25c}
\end{equation}

\section*{Acknowledgement}

We would like to thank former and current members of the IEEE 802.22 committee for important suggestions and discussions, in particular Winston Caldwell, Gerald Chouinard, Carlos Cordeiro, Wendong Hu, Saishankar Nandagopalan, Steve Shellhammer, Carl R. Stevenson, Jianfeng Wang and Lei Zhongding. We would also like to thank Jasper Goseling and Jos Weber from TU Delft for helpful discussions.


\begin{thebibliography}{10}
\providecommand{\url}[1]{#1}
\csname url@rmstyle\endcsname
\providecommand{\newblock}{\relax}
\providecommand{\bibinfo}[2]{#2}
\providecommand\BIBentrySTDinterwordspacing{\spaceskip=0pt\relax}
\providecommand\BIBentryALTinterwordstretchfactor{4}
\providecommand\BIBentryALTinterwordspacing{\spaceskip=\fontdimen2\font plus
\BIBentryALTinterwordstretchfactor\fontdimen3\font minus
  \fontdimen4\font\relax}
\providecommand\BIBforeignlanguage[2]{{%
\expandafter\ifx\csname l@#1\endcsname\relax
\else
\language=\csname l@#1\endcsname
\fi
#2}}

\bibitem{park_submitted_2010_gc}
J.~{Park}, P.~{Pawe{\l}czak}, P.~{Gr{\o}nsund}, and D.~{\v{C}abri{\'c}},
  ``Performance of opportunistic spectrum {OFDMA} network with users of
  different priorities and traffic characteristics,'' in \emph{Proc. IEEE
  GLOBECOM}, Miami, FL, USA, Dec. 6--10, 2010.

\bibitem{staple_spectrum_2004}
G.~{Staple} and K.~{Werbach}, ``The end of spectrum scarcity,'' \emph{{IEEE}
  Spectr.}, vol.~41, no.~3, pp. 48--52, Mar. 2004.

\bibitem{Zhao_sigprocmag_2007}
Q.~{Zhao} and B.~M. {Sadler}, ``A survey of dynamic spectrum access: Signal
  processing, networking, and regulatory policy,'' \emph{{IEEE} Signal
  Processing Mag.}, vol.~24, no.~3, pp. 79--89, May 2007.

\bibitem{nuyami_wimax_book}
L.~Nuyami, \emph{{WiMAX}: Technology for Broadband Wireless Access}.\hskip 1em
  plus 0.5em minus 0.4em\relax Hoboken, NJ, USA: Wiley, 2007.

\bibitem{ieee80220}
\emph{Draft Standard for Local and Metropolitan Area Networks: Standard Air
  Interface for Mobile Broadband Wireless Access Systems Supporting Vehicular
  Mobility: Physical and Media Access Control Layer Specification}, IEEE Std.
  P802.20 Draft 3.0, Mar. 2007.

\bibitem{astely_commag_2009}
D.~{Ast\'{e}ly}, E.~{Dahlman}, A.~{Furusk\"{a}r}, Y.~{Jading},
  M.~{Lindstr\"{o}m}, and S.~{Parkvall}, ``{LTE}: The evolution of mobile
  broadband,'' \emph{{IEEE} Commun. Mag.}, vol.~47, no.~4, pp. 44--51, Apr.
  2009.

\bibitem{weiss_commag_2004}
T.~A. {Weiss} and F.~K. {Jondral}, ``Spectrum pooling: an innovative strategy
  for the enhancement of spectrum efficiency,'' \emph{{IEEE} Commun. Mag.},
  vol.~42, no.~3, pp. S8--S14, Mar. 2004.

\bibitem{mahmoud_wcom_2009}
H.~A. {Mahmoud}, T.~{Y\u{u}cek}, and H.~{Arslan}, ``{OFDM} for cognitive radio:
  merits and challenges,'' \emph{{IEEE} Wireless Commun. Mag.}, vol.~16, no.~2,
  pp. 6--15, Jan./Feb. 2009.

\bibitem{ieee80222}
\emph{Draft Standard for Wireless Regional Area Networks Part 22: Cognitive
  Wireless {RAN} Medium Access Control ({MAC}) and Physical Layer ({PHY})
  Specifications: Policies and Procedures for Operation in the {TV} Bands},
  IEEE Std. P802.22 Draft 7.0, Dec. 2010.

\bibitem{cordeiro_book10}
C.~{Cordeiro}, D.~{Cavalcanti}, and S.~{Nandagopalan}, ``Cognitive radio for
  broadband wireless access in {TV} bands: The {IEEE} 802.22 standard,'' in
  \emph{Cognitive Radio Communications and Networks: Principles and Practice},
  A.~M. {Wyglinski}, M.~{Nekovee}, and Y.~T. {Hou}, Eds.\hskip 1em plus 0.5em
  minus 0.4em\relax Amsterdam, The Netherlands: Elsevier, Inc., 2009.

\bibitem{stevenson_commag09}
C.~R. {Stevenson}, G.~{Chouinard}, Z.~{Lei}, W.~{Hu}, S.~J. {Shellhammer}, and
  W.~{Caldwell}, ``{IEEE 802.22}: The first cognitive radio wireless regional
  area network standard,'' \emph{{IEEE} Commun. Mag.}, vol.~47, no.~1, pp.
  130--138, Jan. 2009.

\bibitem{wang_dyspan_2010}
J.~{Wang}, M.~S. {Song}, S.~{Santhiveeran}, K.~{Lim}, G.~{Ko}, K.~{Kim}, and
  S.~H. {Hwang}, ``First cognitive radio networking standard for
  personal/portable devices in {TV} white spaces,'' in \emph{Proc. IEEE
  DySPAN}, Singapore, Apr. 6--9, 2010.

\bibitem{ieee80211af}
\emph{Draft Standard for Information Technology; Telecommunications and
  information exchange between systems; Local and metropolitan area networks;
  Specific Requirements; Part 11: Wireless {LAN} Medium Access Control ({MAC})
  and Physical Layer ({PHY}) specifications; Amendment 5: {TV} White Spaces
  Operation}, IEEE Std. P802.11af Draft 1.0, Jan. 2011.

\bibitem{durresi_tranbroad_07}
A.~{Durresi} and V.~{Paruchuri}, ``Broadcast protocol for energy-constrained
  networks,'' \emph{{IEEE} Trans. Broadcast.}, vol.~53, no.~1, pp. 112--119,
  Mar. 2007.

\bibitem{liang_wcommag08}
Y.-C. {Liang}, A.~T. {Hoang}, and H.-H. {Chen}, ``Cognitive radio on {TV}
  bands: A new approach to prove wireless connectivity for rural areas,''
  \emph{{IEEE} Wireless Commun. Mag.}, vol.~15, no.~3, pp. 16--22, June 2008.

\bibitem{leu_procieee_2009}
A.~E. {Leu}, B.~L. {Mark}, and M.~A. {McHenry}, ``A framework for cognitive
  {WiMAX} frequency agility,'' \emph{Proc. {IEEE}}, vol.~97, no.~4, pp.
  755--773, Apr. 2009.

\bibitem{song_chinacom_2008}
L.~{Song}, C.~{Feng}, Z.~{Zeng}, and X.~{Zhang}, ``Research on {WRAN} system
  level simulation platform design,'' in \emph{Proc. ICST Chinacom}, Hangzhou,
  China, Aug. 25--27, 2008.

\bibitem{chen_globecom_2007}
H.-S. {Chen}, W.~{Gao}, and D.~G. {Daut}, ``Spectrum sensing using
  cyclostationary properties and application to {IEEE} 802.22 {WRAN},'' in
  \emph{Proc. IEEE GLOBECOM}, Washington, DC, USA, Nov. 26--30, 2007.

\bibitem{chen_icc_2007}
------, ``Signature based spectrum sensing algorithms for {IEEE} 802.22
  {WRAN},'' in \emph{Proc. IEEE ICC}, Glasgow, Scotland, UK, EU, June 24--28,
  2007.

\bibitem{Park_jssc_09}
J.~{Park}, T.~{Song}, J.~{Hur}, S.~M. {Lee}, J.~{Choi}, K.~{Kim}, K.~{Lim},
  C.-H. {Lee}, H.~{Kim}, and J.~{Laskar}, ``A fully integrated {UHF}-band
  {CMOS} receiver with multi-resolution spectrum sensing ({MRSS}) functionality
  for {IEEE} 802.22 cognitive radio applications,'' \emph{{IEEE} J. Solid-State
  Circuits}, vol.~44, no.~1, pp. 258--268, Jan. 2009.

\bibitem{kim_casii_2008}
H.~{Kim}, J.~{Kim}, S.~{Yang}, M.~{Hong}, and Y.~{Shin}, ``An effective
  {MIMO-OFDM} system for {IEEE} 802.22 {WRAN} channels,'' \emph{{IEEE} Trans.
  Circuits Syst. {II}}, vol.~55, no.~8, pp. 821--825, Aug. 2008.

\bibitem{niyato_ieeewcm_2009}
D.~{Niyato}, E.~{Hossain}, and Z.~{Han}, ``Dynamic spectrum access in {IEEE}
  802.22-based cognitive wireless networks: A game theoretic model for
  competitive spectrum bidding and pricing,'' \emph{{IEEE} Wireless Commun.
  Mag.}, vol.~16, no.~2, pp. 16--23, Apr. 2009.

\bibitem{al-zubi_globecom_2009}
R.~{Al-Zubi}, M.~Z. {Siam}, and M.~{Krunz}, ``Coexistence problem in {IEEE}
  802.22 wireless regional area networks,'' in \emph{Proc. IEEE GLOBECOM},
  Honolulu, HI, USA, Nov. 30~--~Dec. 4, 2009.

\bibitem{huang_icics_2009}
D.~{Huang}, C.~{Miao}, Y.~{Miao}, and Z.~{Shen}, ``A game theory approach for
  self-coexistence analysis among {IEEE} 802.22 networks,'' in \emph{Proc. IEEE
  ICICS}, Macau, China, Dec. 7--10, 2009.

\bibitem{sengupta_globecom_2008}
S.~{Sengupta}, R.~{Chandramouli}, S.~{Brahma}, and M.~{Chatterjee}, ``A game
  theoretic framework for distributed self-coexistence among {IEEE} 802.22
  networks,'' in \emph{Proc. IEEE GLOBECOM}, New Orleans, LA, USA, Nov.
  30~--~Dec. 4, 2008.

\bibitem{kim_infocom_2010}
H.~{Kim} and K.~G. {Shin}, ``Asymmetry-aware real-time distributed joint
  resource allocation in ieee 802.22 {WRANs},'' in \emph{Proc. IEEE INFOCOM},
  San Diego, CA, USA, Mar. 15--19, 2010.

\bibitem{ko_tvt_2010}
C.-H. {Ko} and H.-Y. {Wei}, ``Game theoretical resource allocation for
  inter-{BS} coexistence in {IEEE} 802.22,'' \emph{{IEEE} Trans. Veh.
  Technol.}, vol.~59, no.~4, pp. 1729--1744, May 2010.

\bibitem{hu_commmag07}
W.~{Hu}, D.~{Willkomm}, M.~{Abusubiah}, J.~{Gross}, G.~{Vlantis}, M.~{Gerla},
  and A.~{Wolisz}, ``Dynamic frequency hopping communities for efficient {IEEE
  802.22} operation,'' \emph{{IEEE} Commun. Mag.}, vol.~45, no.~5, pp. 80--86,
  May 2007.

\bibitem{segupta_iccworkshop_2009}
S.~{Segupta}, M.~{Chatterje}, and R.~{Chandramouli}, ``A coordinated
  distributed scheme for cognitive radio based {IEEE} 802.22 wireless mesh
  networks,'' in \emph{Proc. IEEE CogNet (IEEE ICC 2008 Workshop)}, Beijing,
  China, May 19--23, 2008.

\bibitem{elayoubi_ton08}
S.-E. {Elayoubi} and B.~{Fouresit\'e}, ``Performance evaluation of admission
  control and adaptive modulation in {OFDMA} {WiMax} systems,''
  \emph{{IEEE/ACM} Trans. Networking}, vol.~16, no.~5, pp. 1200--1211, Oct.
  2008.

\bibitem{luo_twc_2008}
T.~{Luo}, W.~{Xiang}, and H.-H. {Chen}, ``A subcarrier allocation scheme for
  cognitive radio systems based on multi-carrier modulation,'' \emph{{IEEE}
  Trans. Wireless Commun.}, vol.~7, no.~9, pp. 3335--3340, Sept. 2008.

\bibitem{Xing_jsac_2006}
Y.~{Xing}, R.~{Chandramouli}, S.~{Mangold}, and S.~{Shankar N}, ``Dynamic
  spectrum access in open spectrum wireless networks,'' \emph{{IEEE} J. Select.
  Areas Commun.}, vol.~24, no.~3, pp. 626--637, Mar. 2006.

\bibitem{pawelczak_tvt_2009}
P.~Pawe{\l}czak, S.~{Pollin}, H.-S.~W. {So}, A.~{Bahai}, R.~V. {Prasad}, and
  R.~{Hekmat}, ``Performance analysis of multichannel medium access control
  algorithms for opportunistic spectrum access,'' \emph{{IEEE} Trans. Veh.
  Technol.}, vol.~58, no.~6, pp. 3014--3031, July 2009.

\bibitem{Chou_jsac_2007}
C.-T. {Chou}, S.~{Shankar N}, H.~{Kim}, and K.~G. {Shin}, ``What and how much
  to gain by spectrum agility?'' \emph{{IEEE} J. Select. Areas Commun.},
  vol.~25, no.~3, pp. 576--588, Apr. 2007.

\bibitem{Park_arxiv_2009}
\BIBentryALTinterwordspacing
J.~Park, P.~Pawe{\l}czak, and D.~\v{C}abri\'{c}. (2010) Performance of joint
  spectrum sensing and {MAC} algorithms for multichannel opportunistic spectrum
  access ad hoc networks. Accepted for publication, {IEEE} Trans. Mobile
  Comput. [Online]. Available: \url{http://arxiv.org/abs/0910.4704}
\BIBentrySTDinterwordspacing

\bibitem{Liang_twc_2008}
Y.-C. {Liang}, Y.~{Zeng}, E.~C. {Peh}, and A.~T. {Hoang}, ``Sensing throughput
  tradeoff in cognitive radio networks,'' \emph{{IEEE} Trans. Wireless
  Commun.}, vol.~7, no.~4, pp. 1326--1337, Apr. 2008.

\bibitem{Hoang_twc_2009}
A.~T. {Hoang}, Y.-C. {Liang}, D.~T.~C. {Wong}, Y.~{Zeng}, and R.~{Zhang},
  ``Opportunistic spectrum access for energy-constrained cognitive radios,''
  \emph{{IEEE} Trans. Wireless Commun.}, vol.~8, no.~3, pp. 1206--1211, Mar.
  2009.

\bibitem{Papadimitratos_commag_2005}
P.~{Papadimitratos}, S.~{Sankaranarayanan}, and A.~{Mishra}, ``A bandwidth
  sharing approach to improve licensed spectrum utilization,'' \emph{{IEEE}
  Commun. Mag.}, vol.~43, no.~12, pp. S10--S14, Dec. 2005.

\bibitem{gronsund_pimrc_2009}
P.~{Gr{\o}nsund}, H.~N. {Pham}, and P.~E. {Engelstad}, ``Towards dynamic
  spectrum access in primary {OFDMA} systems,'' in \emph{Proc. IEEE PIMRC},
  Tokyo, Japan, Sept. 13--16, 2009.

\bibitem{zhang_tvt_2011}
Y.~{Zhang} and C.~{Leung}, ``A distributed algorithm for resource allocation in
  {OFDM} cognitive radio systems,'' \emph{{IEEE} Trans. Veh. Technol.},
  vol.~60, no.~2, pp. 546--554, Feb. 2011.

\bibitem{zhang_tvt_2009}
------, ``Cross-layer resource allocation for mixed services in multiuser
  {OFDM}-based cognitive radio systems,'' \emph{{IEEE} Trans. Veh. Technol.},
  vol.~58, no.~8, pp. 4605--4619, Oct. 2009.

\bibitem{cheng_icc_2008}
P.~{Cheng}, Z.~{Zhang}, H.~{Huang}, and P.~{Qiu}, ``A distributed algorithm for
  optimal resource allocation in cognitive {OFDMA} systems,'' in \emph{Proc.
  IEEE ICC}, Beijing, China, May 19--23, 2008.

\bibitem{zhang_comml_2009}
Y.~{Zhang} and C.~{Leung}, ``Resource allocation for non-real-time services in
  {OFDM}-based cognitive radio systems,'' \emph{{IEEE} Commun. Lett.}, vol.~13,
  no.~1, pp. 16--18, Jan. 2009.

\bibitem{Gerihofer_commag_2007}
S.~{Geirhofer}, L.~{Tong}, and B.~M. {Sandler}, ``Dynamic spectrum access in
  the time domain: Modeling and exploiting white space,'' \emph{{IEEE} Commun.
  Mag.}, vol.~45, no.~5, pp. 66--72, May 2007.

\bibitem{Huang_infocom_2009}
S.~{Huang}, X.~{Liu}, and Z.~{Ding}, ``Optimal sensing-transmission structure
  for dynamic spectrum access,'' in \emph{Proc. IEEE INFOCOM}, Rio De Janeiro,
  Brazil, Apr. 19--25, 2009.

\bibitem{gambini_twc_2008}
J.~{Gambini}, O.~{Simeone}, Y.~{Bar-Ness}, U.~{Spagnolini}, and T.~{Yu},
  ``Packet-wise vertical handover for unlicensed multi-standard spectrum access
  with cognitive radios,'' \emph{{IEEE} Trans. Wireless Commun.}, vol.~7,
  no.~12, pp. 5172--5176, Dec. 2008.

\bibitem{vanmieghem_book_2006}
P.~{Van Mieghem}, \emph{Performance Analysis of Communications Networks and
  Systems}.\hskip 1em plus 0.5em minus 0.4em\relax Cambridge University Press,
  2006.

\bibitem{wellens_phycom_2009}
M.~{Wellens}, J.~{Riihij{\"a}rvi}, and P.~{M{\"a}h{\"o}nen}, ``Empirical time
  and frequency domain models of spectrum use,'' \emph{Elsevier Physical
  Communication Journal}, vol.~2, no. 1--2, pp. 10--32, Mar.--Jun. 2009.

\bibitem{peh_tvt_2009}
E.~C.~Y. {Peh}, Y.-C. {Liang}, Y.~L. {Guan}, and Y.~{Zeng}, ``Optimization of
  cooperative sensing in cognitive radio networks: A sensing-throughput
  tradeoff view,'' \emph{{IEEE} Trans. Veh. Technol.}, vol.~58, no.~9, pp.
  5294--5299, Nov. 2009.

\bibitem{song_twc_2010}
S.~H. {Song}, K.~{Hamdi}, and K.~B. {Letaief}, ``Spectrum sensing with active
  cognitive systems,'' \emph{{IEEE} Trans. Wireless Commun.}, vol.~9, no.~6,
  pp. 1849--1854, June 2010.

\bibitem{luo_twc_2009}
L.~{Luo}, N.~M. {Neihart}, S.~{Roy}, and D.~J. {Allstot}, ``A two-stage sensing
  technique for dynamic spectrum access,'' \emph{{IEEE} Trans. Wireless
  Commun.}, vol.~8, no.~6, pp. 3028--3037, June 2009.

\bibitem{jeon_tvt_2010}
W.~S. {Jeon} and D.~G. {Jeong}, ``An advanced quiet-period management scheme
  for cognitive radio systems,'' \emph{{IEEE} Trans. Veh. Technol.}, vol.~59,
  no.~3, pp. 1242--1256, Mar. 2010.

\bibitem{Gabran_arxiv_2010}
\BIBentryALTinterwordspacing
W.~{Gabran}, P.~Pawe{\l}czak, and D.~\v{C}abri\'{c}. Throughput and collision
  analysis for multi-channel multi-stage spectrum sensing algorithms. [Online].
  Available: \url{http://arxiv.org/abs/1010.0041}
\BIBentrySTDinterwordspacing

\bibitem{jeon_twc_2008}
W.~S. {Jeon}, D.~G. {Jeong}, J.~A. {Han}, G.~{Ko}, and M.~S. {Song}, ``An
  efficient quiet period management scheme for cognitive radio systems,''
  \emph{{IEEE} Trans. Wireless Commun.}, vol.~7, no.~2, pp. 505--509, Feb.
  2008.

\bibitem{fcc_docket_2008}
{FCC}, ``In the matter of {ET} {Docket} no. 04-18 and {ET} {Docket} no. 02-38
  second report and order and memorandum opinion and order,'' Federal
  Communications Commission, Tech. Rep. 08-260, Nov. 14, 2008.

\bibitem{ngo_tvt_2010}
D.~T. {Ngo}, C.~{Tellambura}, and H.~H. {Nguyen}, ``Resource allocation for
  {OFDMA}-based cognitive radio multicast networks with primary user activity
  consideration,'' \emph{{IEEE} Trans. Veh. Technol.}, vol.~59, no.~4, pp.
  1668--1679, May 2010.

\bibitem{qu_tvt_2010}
D.~{Qu}, Z.~{Wang}, and T.~{Jiang}, ``Extended active interference cancellation
  for sidelobe suppression in cognitive radio {OFDM} systems with cyclic
  prefix,'' \emph{{IEEE} Trans. Veh. Technol.}, vol.~59, no.~4, pp. 1689--1695,
  May 2010.

\bibitem{yuan_tvt_2010}
Z.~{Yuan} and A.~M. {Wyglinski}, ``On sidelobe suppression for
  multicarrier-based transmission in dynamic spectrum access networks,''
  \emph{{IEEE} Trans. Veh. Technol.}, vol.~59, no.~4, pp. 1998--2006, May 2010.

\bibitem{sutton_dyspan_2010}
P.~{Sutton}, B.~{\"{O}zg\"{u}l}, I.~{Macaluso}, and L.~{Doyle}, ``{OFDM}
  pulse-shaped waveforms for dynamic spectrum access,'' in \emph{Proc. IEEE
  DySPAN}, Singapore, Apr. 6--9, 2010.

\bibitem{sharma_jsac2004}
S.~{Sharma}, N.~{Zhu}, and T.-c. {Chiueh}, ``Low-latency mobile {IP} handoff
  for infrastructure-mode wireless {LAN},'' \emph{{IEEE} J. Select. Areas
  Commun.}, vol.~22, no.~4, pp. 643--652, May 2004.

\bibitem{raniwala_infocom_2005}
A.~{Raniwala} and T.-c. {Chiueh}, ``Architecture and algorithms for an {IEEE}
  802.11-based multi-channel wireless mesh network,'' in \emph{Proc. IEEE
  INFOCOM}, Miami, FL, USA, Mar. 12--17, 2005.

\bibitem{reihl_report_2006}
E.~{Reihl}, ``Wireless microphone characteristics,'' IEEE, Tech. Rep.
  802.22-06/0070r0, May 2006.

\bibitem{lam_commag_1997}
D.~{Lam}, D.~C. {Cox}, and J.~{Widom}, ``Teletraffic modeling for personal
  communication services,'' \emph{{IEEE} Commun. Mag.}, vol.~35, no.~2, pp.
  79--87, Feb. 1997.

\bibitem{etsitr102546}
\emph{Electromagnetic Compatibility and Radio Spectrum Matters ({ERM});
  Technical Characteristics for Professional Wireless Microphone Systems
  ({PWMS}); System Reference Model}, ETSI Std. TR 102 546 v1.1.1, Apr. 2007.

\bibitem{chen_jsac_2011}
H.-S. {Chen} and W.~{Gao}, ``Spectrum sensing for {TV} white space in north
  america,'' \emph{{IEEE} J. Select. Areas Commun.}, vol.~29, no.~2, pp.
  316--326, Feb. 2011.

\end{thebibliography}
\end{document}